\documentclass[sigconf]{acmart}
\usepackage[utf8]{inputenc}
\usepackage{siunitx}

\AtBeginDocument{%
  \providecommand\BibTeX{{%
    \normalfont B\kern-0.5em{\scshape i\kern-0.25em b}\kern-0.8em\TeX}}}

\makeatletter
\newenvironment{tapscallout}{%
  \def\FrameCommand{\fboxrule=0.5pt \fboxsep=6pt \fcolorbox{black!20}{green!10}}%
  \MakeFramed{\advance\hsize-\width \FrameRestore}%
}{\endMakeFramed}
\makeatother

\makeatletter
\newenvironment{tapsbox}[1][green!10]{%
  \def\FrameCommand{\fboxrule=0.5pt \fboxsep=6pt \fcolorbox{black!20}{#1}}%
  \MakeFramed{\advance\hsize-\width \FrameRestore}%
}{\endMakeFramed}
\makeatother

\usepackage{multirow}
\usepackage{rotating}
\usepackage{colortbl}

\usepackage{graphicx}
\usepackage{subcaption}

\usepackage{framed}

\usepackage{algorithm}
\usepackage{algorithmic}

\usepackage{tikz}
\usepackage{enumitem}

\usetikzlibrary{shapes.geometric, arrows, positioning, fit}

\copyrightyear{2026}
\acmYear{2026}
\setcopyright{cc}
\setcctype{by-nc-nd}
\acmConference[CHI '26]{Proceedings of the 2026 CHI Conference on Human Factors in Computing Systems}{April 13--17, 2026}{Barcelona, Spain}
\acmBooktitle{Proceedings of the 2026 CHI Conference on Human Factors in Computing Systems (CHI '26), April 13--17, 2026, Barcelona, Spain}
\acmPrice{}
\acmDOI{10.1145/3772318.3791270}
\acmISBN{979-8-4007-2278-3/2026/04}

\begin{document}


\title[Co-Designing Gaze Gestures with Users and Experts]{The People's Gaze: Co-Designing and Refining Gaze Gestures with General Users and Gaze Interaction Experts}


\author{Yaxiong Lei}
\email{Yaxiong.Lei@st-andrews.ac.uk}
\orcid{0000-0002-0697-7942}
\affiliation{%
  \institution{University of St Andrews}
  \city{St Andrews}
  \country{UK}
}
\affiliation{%
  \institution{University of Essex}
  \city{Colchester}
  \country{UK}
}

\author{Xinya Gong}
\email{xg31@st-andrews.ac.uk}
\orcid{0009-0005-6414-9351}
\affiliation{%
  \institution{University of St Andrews}
  \city{St Andrews}
  \country{UK}
}

\author{Shijing He}
\email{shijing.he@kcl.ac.uk}
\orcid{0000-0003-3697-0706}
\affiliation{%
 \institution{King's College London}
 \city{London}
 \country{UK}
}

\author{Yafei Wang}
\email{wangyafei@dlmu.edu.cn}
\orcid{0000-0002-8005-1718}
\affiliation{%
 \institution{Dalian Maritime University}
 \city{Dalian}
 \country{China}
}

\author{Mohamed Khamis}
\email{Mohamed.Khamis@glasgow.ac.uk}
\orcid{0000-0001-7051-5200}
\affiliation{%
 \institution{University of Glasgow}
 \city{Glasgow}
 \country{UK}
}
 
\author{Juan Ye}
\email{Juan.Ye@st-andrews.ac.uk}
\orcid{0000-0002-2838-6836}
\affiliation{%
 \institution{University of St Andrews}
  \city{St Andrews}
  \country{UK}
}


\renewcommand{\shortauthors}{Yaxiong Lei et al.}

\begin{abstract}

As eye-tracking becomes increasingly common in modern mobile devices, the potential for hands-free, gaze-based interaction grows, but current gesture sets are largely expert-designed and often misaligned with how users naturally move their eyes. To address this gap, we introduce a two-phase methodology for developing intuitive gaze gestures. First, four co-design workshops with 20 non-expert participants generated 102 initial concepts. Next, four gaze interaction experts reviewed and refined these into a set of 32 gestures. We find that non-experts, after a brief introduction, intuitively anchor gestures in familiar metaphors and develop a compositional grammar; i.e., activation (dwell) + action (gaze gesture or blink), to ensure intentionality and mitigate the classic Midas Touch problem. Experts prioritize gestures that are ergonomically sound, aligned with natural saccades, and reliably distinguishable. The resulting user-grounded, expert-validated gesture set, along with actionable design principles, provides a foundation for developing intuitive, hands-free interfaces for gaze-enabled devices.

\end{abstract}




\begin{teaserfigure}
\includegraphics[width=\textwidth]{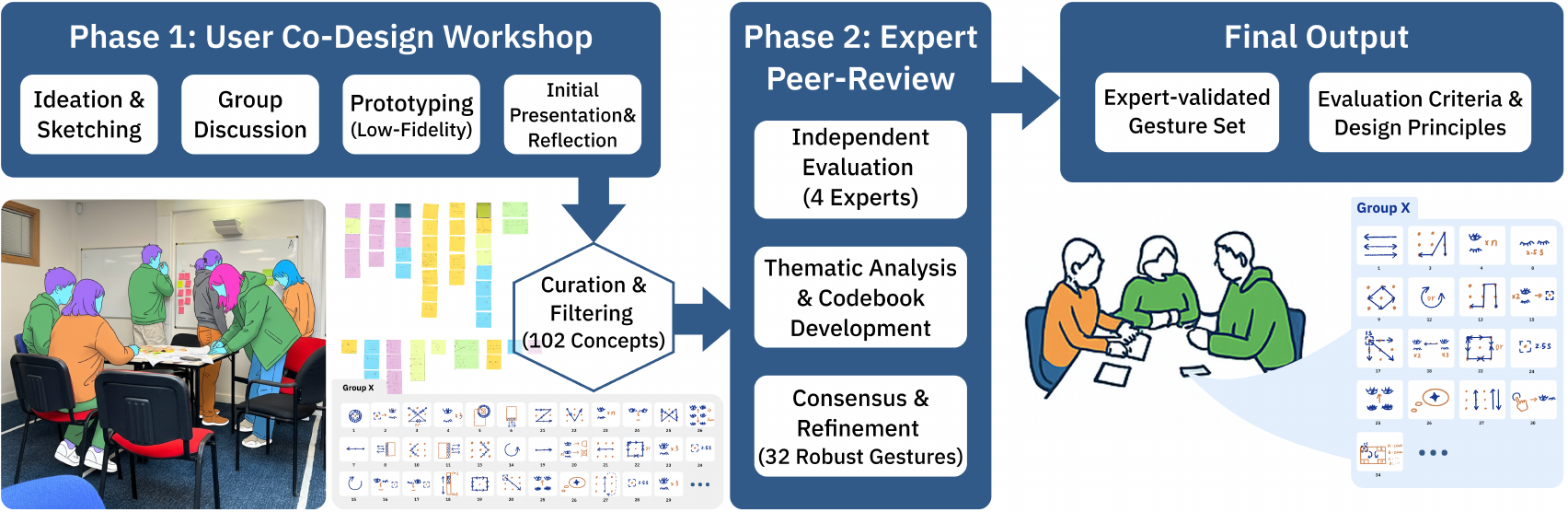}
  \Description{A horizontal workflow diagram illustrating the three stages of the research methodology. The first block is "Phase 1: User Co-Design Workshop," containing steps for "Ideation \& Sketching," "Group Discussion," "Prototyping (Low-Fidelity)," and "Initial Presentation \& Reflection." This leads to a central hexagon labeled "Curation \& Filtering (102 Concepts)." The second block is "Phase 2: Expert Peer-Review," containing steps for "Independent Evaluation (4 Experts)," "Thematic Analysis \& Codebook Development," and "Consensus \& Refinement (32 Gestures)." The final block is "Final Output," listing "Expert-validated Gesture Set" and "Evaluation Criteria \& Design Principles." Illustrations of people in a workshop setting and a grid of gesture icons accompany the text.}
  \caption{Overview of the research process. The Phase 1 involves a co-design workshop with participants, through creative ideation, group discussions, prototyping, and reflection sessions, resulting in 102 initial concepts (reduced to 59 unique gestures after cross-group deduplication). The Phase 2 consists of a structured expert peer-review, including independent evaluation, discussion, thematic analysis, and codebook development, leading to consensus and refinement into 32 gestures. The final output includes the validated gesture set, derived evaluation criteria and design principles.}
  \label{fig:teaser}
\end{teaserfigure}


\maketitle

% ======== Main Content ========%

% \input{section.v2/1-intro.v2}
% \input{section.v2/2-related.v2}
% \input{section.v2/3-methodology.v2}
% \input{section.v2/4-results.v3}
% \input{section.v2/5-discussion.v3}
% \input{section.v2/6-conclusion.v2}

\section{Introduction}
Eye tracking is rapidly transitioning from specialized laboratories to everyday devices. Head-mounted displays such as Apple's \emph{Vision Pro} and Meta's \emph{Quest Pro} now ship with sub-degree eye trackers, while iOS 18 introduces gaze input to mass-market smartphones for the first time~\cite{InthePocket2025VisionPro, MetaHelp2024QuestProEyeTracking, apple2024iphoneeyetracking}. Analysts forecast that the eye-tracking market will more than double in value by 2030, with a compound annual growth rate exceeding 25\% through 2033~\cite{GlobalGrowthInsights2025EyeTrackingMarket, ResearchMarkets2025EyeTrackingCAGR}. Moreover, prior research has demonstrated significant potential, particularly for individuals with motor impairments who rely on gaze as their primary control channel, as well as for able-bodied users seeking hands-free interaction in VR, AR, or mobile scenarios~\cite{severitt2025interplay, fujii2018gaze, krishna2020eye, lei2023end, cheng2024appearance}. Recent research prototypes also push accuracy and low-latency processing by leveraging on-device machine learning and multimodal interaction methods~\cite{lei2023end, majumder2025eye}. Meanwhile, advances in pattern recognition and gaze-estimation algorithms provide the technical foundations for reliable, real-time decoding of gaze gestures (see details in \S \ref{related_gaze_interaction_gestures}). In particular, gaze-estimation pipelines now leverage large-scale datasets and lightweight deep learning backbones, i.e. CNN/MobileViT, to achieve sub-degree angular accuracy on commodity cameras~\cite{krafka2016eye, zhang2015appearance, ververas20243dgazenet}, while transformer-based models drive end-to-end latencies below 30 ms on mobile GPUs~\cite{xu2023fr}.

However, technical capability alone does not ensure usability. Transforming raw eye-tracking data into expressive, discoverable, and user-friendly gestures remains a non-trivial challenge~\cite{zhong2024uncertainty, lei2025quantifying}. Existing gesture vocabularies are often developed by researchers in controlled settings and tested with relatively small, homogeneous user samples. These expert-defined sets may not reflect the mental models or preferences of end-users, especially in everyday mobile contexts. As a result, questions persist around gesture learnability, memorability, and applicability across devices and cultural contexts. Recent improvements in gaze gesture recognition now make it technically feasible to detect and classify gaze gestures in everyday mobile environments (see \S \ref{related_gaze_interaction_gestures}). Yet, the interaction design space remains underdeveloped. We argue that this is a critical juncture for establishing a systematic, user-defined vocabulary of gaze gestures---one that captures natural user behaviours while maintaining ergonomic feasibility and implementation viability. Our study aims to explore the following research questions (RQs):

\begin{enumerate}
  \item \textit{RQ1:} Which gaze gestures do non-expert users design for common mobile commands, and what are their underlying rationales?  
  \item \textit{RQ2:} How are these user-defined gestures assessed by expert evaluators?  
\end{enumerate}

To address these RQs, we conducted a two-phase study to develop and validate a user-defined gaze gesture vocabulary. In Phase 1, we ran four co-design workshops with 20 non-expert participants in the UK, using a 9-point spatial grid~\cite{sanders2014probes} (a 3$\times$3 set of fixed gaze anchor points overlaid on the screen) to ensure gesture feasibility and comparability, to scaffold ideation. Participants generated 102 unique gaze gesture concepts through collaborative sketching, discussion, and reflection. We then removed duplicates and gestures that could not be represented on the 9-point grid, resulting in a finalized set of 59 unique gaze gestures. In Phase 2, four experts in gaze interaction and eye movement research independently reviewed and rated these gestures in a structured peer-review format. This expert input yielded a refined set of 32 gestures deemed feasible, distinct, and ergonomic.

\textbf{Contribution.} First, we present the first empirical insights from participatory co-design workshops, offering a rare bottom-up perspective on how non-experts conceptualize gaze-based interaction on mobile devices. For instance, our results reveal how users anchor gaze gestures in familiar physical metaphors (e.g., tapping or swiping). Second, we contribute a user-defined vocabulary of gaze gestures, systematically derived from over 100 initial ideas and refined through structured expert peer review. This iterative filtering process produces a robust, ergonomic, and distinct set of 32 gaze gestures suitable for real-world use. Third, we identify and formalise a key interaction pattern that we term a \emph{compositional gaze grammar}: an \emph{activate-then-confirm} structure in which an activation primitive (e.g., a short dwell) explicitly ``arms'' the interface and a subsequent confirmation primitive (e.g., a scanpath or blink) commits the command. This grammar adapts direct-manipulation principles of explicit control, feedback, and reversibility~\cite{shneiderman1983direct} to a continuous-signal modality, and provides a reusable design template for separating ``looking'' from ``acting''.
This finding offers a flexible foundation for designing context-aware, hands-free gaze interactions. Overall, our study contributes actionable design principles and an experience-validated gaze gesture set to inform the next generation of gaze-enabled mobile interfaces.

\section{Related Work}\label{sec:related}

\subsection{Gaze Interaction and Gesture Vocabularies} \label{related_gaze_interaction_gestures}

Calibration drift~\cite{lei2025mac, muller2019reducing}, limited spatial precision~\cite{lei2025quantifying}, and motion-induced input or prediction uncertainty~\cite{zhong2024uncertainty, cai2023source} continue to hinder the deployment of classic point-and-click techniques~\cite{heikkila2009speed}, as even small deviations in tracking accuracy or unstable gaze signals can significantly compromise target selection, slow interaction speed, and increase error rates in real-world conditions. To address these challenges, researchers have proposed a range of gaze interaction methods, which are broadly categorized into three primary modalities~\cite{lei2023end, Khamis2018Survey}. \textbf{Dwell-time} selection, while intuitive, suffers from the well-documented Midas Touch problem of distinguishing intentional fixations from natural viewing behaviour~\cite{jacob1990you, lei2023dynamicread, namnakani2025stretch}. \textbf{Pursuit-based} methods leverage smooth eye movements that follow dynamic targets, showing particular promise in mobile contexts~\cite{Namnakani2023CompareGaze}. Finally, \textbf{Gaze gestures}, deliberate eye movement patterns used to encode commands, offer distinct advantages by relying on relative motion rather than absolute positioning, thereby mitigating calibration drift and accuracy limitations~\cite{lei2023dynamicread, muller2019reducing, Khamis2018Survey, lei2023end, ghosh2023automatic}. Fundamentally, these alternative methods manage constraints of the oculomotor system and human motor control: the eyes are optimised for rapid, ballistic saccades rather than smooth, precise tracing~\cite{kandell2021principles, duong2008neuro}, and gaze must be continuously allocated between task demands and stimulus-driven capture~\cite{land2009vision, jacob1990you}. These links to motor control and cognitive load theories resurface in our experts' ergonomic critiques in \S\ref{result_phase_2}.
On the pattern recognition side, elastic time-series techniques such as \emph{Dynamic Time Warping}~\cite{sakoe1978dynamic} and probabilistic sequence models (e.g., HMMs) have been complemented by deep convolutional recurrent networks that learn spatio-temporal features end-to-end~\cite{mohammadi2024deep}. For 2-D path shapes, robust contour descriptors like \emph{Shape Contexts}~\cite{belongie2002shape} and point-set metrics such as the \emph{Modified Hausdorff Distance}~\cite{dubuisson1994modified} enable rotation- and scale-invariant matching, even under conditions of noise or partial occlusion.

Furthermore, early gaze gesture vocabularies were largely researcher-driven rather than user-driven. Wobbrock et al.~\cite{wobbrock2008longitudinal} pioneered discrete consecutive gestures using the four screen corners as anchors, effectively forming a 4-point grid system that supported alphabet mappings based on letter shapes. Porta and Turina~\cite{porta2008eye} extended this approach to a 9-point grid in their Eye-S system, enabling both command gestures and alphabetic input. More recently, alternative anchoring strategies have been explored: Kim et al.~\cite{kim2022lattice} introduced a 7$\times$7 lattice with 49 visual anchors to improve stability, while Istance et al.~\cite{istance2017supporting} investigated geometric action areas to guide gesture execution. Despite these innovations, most existing vocabularies remain predominantly expert-designed with minimal ordinary user input, i.e.~\cite{jacob1990you, istance2017supporting, drewes2007interacting, bature2023boosted, rozado2015controlling, li2021evaluating, rozado2012gliding, lei2023dynamicread, li2017gazture, chen2019gaze}. A notable exception is Zhang et al.'s GazeSpeak~\cite{Zhang2017GazeSpeak}, which incorporated user feedback for communication in ALS contexts, though its focus was on a specific disability setting rather than general-purpose interaction. This gap highlights the need for participatory approaches to vocabulary development.

\subsection{Participatory Design for Gesture Elicitation} \label{related_participatory_design}

The methodological foundation for user-defined gesture creation rests on Wobbrock et al.'s seminal elicitation framework~\cite{wobbrock2009user}, which inverts traditional design by presenting desired effects before soliciting gestural causes. This ``effect-first, cause-second'' approach has demonstrated consistent superiority over expert-designed gestures across multiple domains. Villarreal-Narvaez et al.'s meta-analysis of 216 hand and body gesture elicitation studies involving 148,340 gestures from 5,458 participants establishes robust evidence for participatory approaches, identifying critical factors including participant diversity, referent presentation methods, and environmental context~\cite{Villarreal2020GestureElicitationReview}. For gaze-specific contexts, Delamare et al.~\cite{delamare2017designing} adapt participatory methods through their G3 system, structuring gaze gesture design around three fundamental phases: Exploration (discovering available gestures), Guidance (executing with visual support), and Return (completing the gesture). Their work highlights unique considerations for gaze gesture elicitation, including eye fatigue, social acceptability, and the challenge of providing adequate visual guidance without interfering with natural eye movements. Recent methodological advances emphasise two-phase approaches combining initial elicitation with systematic evaluation~\cite{vogiatzidakis2018gesture}. Troiano et al.~\cite{troiano2014user} structured this through function identification, gesture elicitation, and evaluation of memorability, comfort, and semantic matching. The Guessability + Evaluation framework~\cite{danielescu2022iterative} further refined this methodology, with Phase 1 focusing on diverse user elicitation and Phase 2 emphasising performance validation with both original and new participants. These two-phase methodologies directly inform our study design, though require adaptation for gaze-specific constraints and cross-platform deployment.

\subsection{Research Gaps and Positioning} \label{related_research_gaps}

Despite rapid technical advances in mobile-compatible eye-tracking systems and real-time gaze estimation pipelines, several critical gaps persist in the design of gaze gesture vocabularies. First, most existing gesture sets are predominantly designed by researchers or domain experts, with \textbf{minimal participation from end-users in shaping the gaze gestures}. This top-down approach risks overlooking intuitive, memorable, and culturally grounded gestures that emerge from everyday user experience. As our findings show, non-expert participants often anchor gaze gestures in familiar metaphors drawn from touch interaction (e.g., tapping, swiping), which are rarely reflected in expert-defined vocabularies. Second, current research rarely addresses the unique constraints of mobile settings, where users must interact while on the move, in varied lighting conditions, or without fixed head orientation. Gesture sets validated in stationary VR or desktop contexts may not translate well to mobile devices. This creates a gap in vocabularies optimized for mobile ergonomics, contextual constraints, and spontaneous usage. Third, prior work often lacks ecological validity. Evaluations are frequently confined to laboratory conditions with limited consideration of situational diversity such as physical movement, multitasking scenarios, or the social acceptability of performing gaze gestures in public. Without addressing these real-world variables, gesture vocabularies risk being technically impressive but practically unusable.

To address these gaps, we employ a two-phase methodology that explicitly couples participatory design with expert evaluation. Phase~1 engages 20 non-expert participants in co-design workshops to elicit a bottom-up vocabulary of 102 initial concepts. Phase~2 applies structured peer review by four gaze interaction experts to filter and refine these into a curated set of 32 gestures. Unlike prior studies on gaze gestures, the design of these gestures focuses more on enabling devices to recognise gestures more effectively and accurately~\cite{wobbrock2008longitudinal, porta2008eye, li2017gazture, Chen2021GazeGestureGCN, bature2023boosted, li2021evaluating, rozado2012gliding, chen2019gaze}, this combined approach offers both methodological rigour and ecological relevance, producing a gesture vocabulary that is grounded in user expectations, validated by expert experience scrutiny, and viable for deployment on gaze-enabled mobile platforms.

\section{Methodology}

\subsection{Participant Recruitment} \label{method_recruitment}
We began phase 1 participant recruitment through social media outreach (e.g., Facebook) and mailing lists of both students and staff at the first author's university by distributing a recruitment advertisement that included a screening survey. All participants were required to meet the following criteria: (1) aged 18 years or older; (2) sufficient English proficiency to understand study instructions and complete tasks during co-design sessions. A total of 20 participants were recruited, divided into four groups (A, B, C, D) of five participants each (see Phase 1 participants' demographic in Table \ref{app:participants}).  
Familiarity with gaze interaction had a mean score of 1.65 ($SD = 0.88$). The average workshop duration was 66.3 minutes, and each participant received a £20 Amazon voucher as compensation.

After completing the phase 1 study, we recruited a panel of four experts in gaze interaction through academic research communities for phase 2. Selection criteria required that participants (1) had published at least four peer-reviewed papers on gaze-based interaction and/or eye-movement research; (2) had direct experience conducting user research; and (3) possessed knowledge of eye-tracking technologies and devices across mobile platforms. The final panel consisted of three males and one female, with an average age of 36.25 years ($SD = 5.13$). These experts participated in a structured evaluation of the gestures proposed in phase 1, and each was compensated with a £20 Amazon voucher for their time and contributions. All expert participants' demographic information can be found in Table \ref{app:expert-demographics}.

\renewcommand{\arraystretch}{1}
\begin{table*}[!ht]
 \centering
 \fontsize{7pt}{8pt}\selectfont
 \resizebox{1\textwidth}{!}{%
 \begin{tabular}{llllllc}
 \toprule 
 \textbf{ID} & \textbf{Age} & \textbf{Gender} & \textbf{Education} & \textbf{Ethnicity} & \textbf{Background | Role} & \textbf{Familiarity} \\ 
 \midrule
 S1-GA-P1 & 25--34 & Female & Master's degree & Asian & Business Management | University administration & 1 \\
 S1-GA-P2 & 35--44 & Male & Associate/Technical degree & Asian & N/A | HR & 1 \\
 S1-GA-P3 & 18--24 & Male & Bachelor's degree & White & Sociology | Student & 1 \\
 S1-GA-P4 & 25--34 & Male & Bachelor's degree & White & Computer Science | Student & 2 \\
 S1-GA-P5 & 45--54 & Male & High school or equivalent & Asian & N/A | Maintenance worker & 1 \\
 S1-GB-P6 & 35--44 & Female & Associate/Technical degree & Black & N/A | Sales & 1 \\
 S1-GB-P7 & 25--34 & Female & Master's degree & Asian & Education | Teaching assistant & 2 \\
 S1-GB-P8 & 25--34 & Female & Doctoral degree & White & Computer Science | Student & 4 \\
 S1-GB-P9 & 18--24 & Male & Bachelor's degree & Black & Biological Sciences | Student & 2 \\
 S1-GB-P10 & 25--34 & Female & Associate/Technical degree & White & N/A | Delivery & 1 \\
 S1-GC-P11 & 35--44 & Female & Master's degree & White & Fine Art | Freelancer & 1 \\
 S1-GC-P12 & 18--24 & Female & Bachelor's degree & Black & Digital Media | Student & 3 \\
 S1-GC-P13 & 25--34 & Male & High school or equivalent & Asian & N/A | Unemployed & 1 \\
 S1-GC-P14 & 25--34 & Male & High school or equivalent & White & N/A | Track driver & 1 \\
 S1-GC-P15 & 18--24 & Female & Master's degree & Asian & Software Engineering | Student & 3 \\
 S1-GD-P16 & 45--54 & Female & High school or equivalent & White & N/A | Unemployed & 2 \\
 S1-GD-P17 & 18--24 & Male & Bachelor's degree & Asian & Marketing | Marketing & 1 \\
 S1-GD-P18 & 18--24 & Female & Bachelor's degree & White & Chemistry Engineering | Student & 2 \\
 S1-GD-P19 & 18--24 & Female & Bachelor's degree & White & Gender Study | Student & 2 \\
 S1-GD-P20 & 25--34 & Male & Associate/Technical degree & Asian & N/A | Sales & 1 \\
\bottomrule
\end{tabular}}
 \caption{Phase 1 Participant Demographics ($N=20$) and Gaze Interaction Familiarity (1--5).}
 \label{app:participants}
\end{table*}

% %\begin{table*}[ht]
% % \centering
% % \caption{Aggregated demographic of the 60 participants in this study. \textcolor{red}{TODO: @Yaxiong Update this demographic info!}}
% % \label{tab:aggregated_demo_data}
% % \fontsize{8pt}{8pt}\selectfont
% % \resizebox{1\textwidth}{!}{
% % \begin{tabular}{llllllll}
% % \toprule
% % \textbf{Gender} & \textbf{Ethnicity} & \textbf{Age} & \textbf{Education} & \textbf{Wear glass} & \textbf{Vision condition} & \textbf{Height (cm)} & \textbf{Face ID in smartphone} %& \textbf{Eye-tracking AR/VR headsets}
% % \\
% % \midrule
% % Woman (33) & White (21) & 18-24 (29) & Bachelor (26) & Yes (32) & None/No known (31) & 160–169 (19) & Yes (60) %& Yes (11)
%  \\
%  Man (24) & Asian (19) & 25-34 (16) & Master (19) & No (28) & Myopia (25) & 170–179 (17) & %& No (49) 
%  \\
%  Non-binary (3) & Black (14) & 35-44 (8) & Doctoral (10) & & Astigmatism (9) & 180–189 (12)
%  \\
%  & Mixed (4) & 45-54 (5) & High school (5) & & & 150–159 (8)
%  \\
%  & Other (2) & $\geq$ 55 (2) & & & & 190+ (4) 
%  \\
%  \bottomrule
%  \end{tabular}}
% \end{table*}

\begin{table*}[!ht]
 \centering
 \fontsize{7pt}{8pt}\selectfont
 \resizebox{\textwidth}{!}{%
 \begin{tabular}{llllllll}
 \toprule 
 \textbf{ID} & \textbf{Age} & \textbf{Gender} & \textbf{Education} & \textbf{Ethnicity} & \textbf{Background | Role} & \textbf{Expertise Area} \\ 
 \midrule
 S2-EP1 & 25--34 & Female & Doctor's degree & Asian & Computer Science | Faculty  & Human Computer Interaction | Gaze Interaction  \\
 S2-EP2 & 35--44 & Male & Doctor's degree & Asian & Computer Science | Faculty & Gaze Estimation | Gaze Interaction | Ubiquitous computing \\
 S2-EP3 & 35--44 & Male & Doctor's degree & White & Computer Science | Faculty & Human Computer Interaction | User-Centred Design  \\
 S2-EP4 & 25--34 & Male & Doctor's degree & White & Computer Science | Faculty & Gaze Estimation | Eye Tracking | Eye Movements\\
\bottomrule
\end{tabular}}
 \caption{Phase 2 Expert Demographics ($N=4$). Faculty includes: postdoctoral researchers, assistant to associate/full professors.}
 \label{app:expert-demographics}
\end{table*}

\subsection{Phase 1 Procedure} \label{method_phase1_procedure}

Following established elicitation methods~\cite{UserInterviewsQualitativeSampleSizes, wobbrock2009user} and best practices~\cite{vogiatzidakis2018gesture}, participants designed gestures for discrete and continuous actions. They were prompted with common commands (e.g., \textit{open}, \textit{return}, \textit{copy}, \textit{zoom in/out}, \textit{light up/down}, etc.) to inspire their designs and continued generating ideas until saturation was reached. We selected these target commands to balance ecological relevance and ease of elicitation: they are commonplace operations in both mobile touchscreen and desktop environments, which allowed participants to quickly understand the scenarios without domain-specific training. Participants were not instructed to prioritise any particular commands; and were free to decide how to position and arrange the gesture designs in sketches. To support creativity while ensuring feasibility, we introduced a 9-point spatial grid~\cite{sanders2014probes}, which standardized gaze paths, lowered entry barriers, and filtered impractical proposals. Four workshops were run in parallel to enable cross-scoring. Prior to the session, participants received a short packet introducing gaze interaction, design principles (simplicity, distinctiveness, comfort, memorability), and technical constraints (jitter, calibration drift, hardware variability)~\cite{Muller2007ParticipatoryDesignThirdSpace}. Each session generated 12--24 gesture candidates with sketches, labels, rationales, and platform tags, supplemented by think-aloud notes and discussions. The process comprised four steps (Fig.~\ref{fig:phase1_process}):

\begin{figure}
    \centering
    \includegraphics[width=0.99\linewidth]{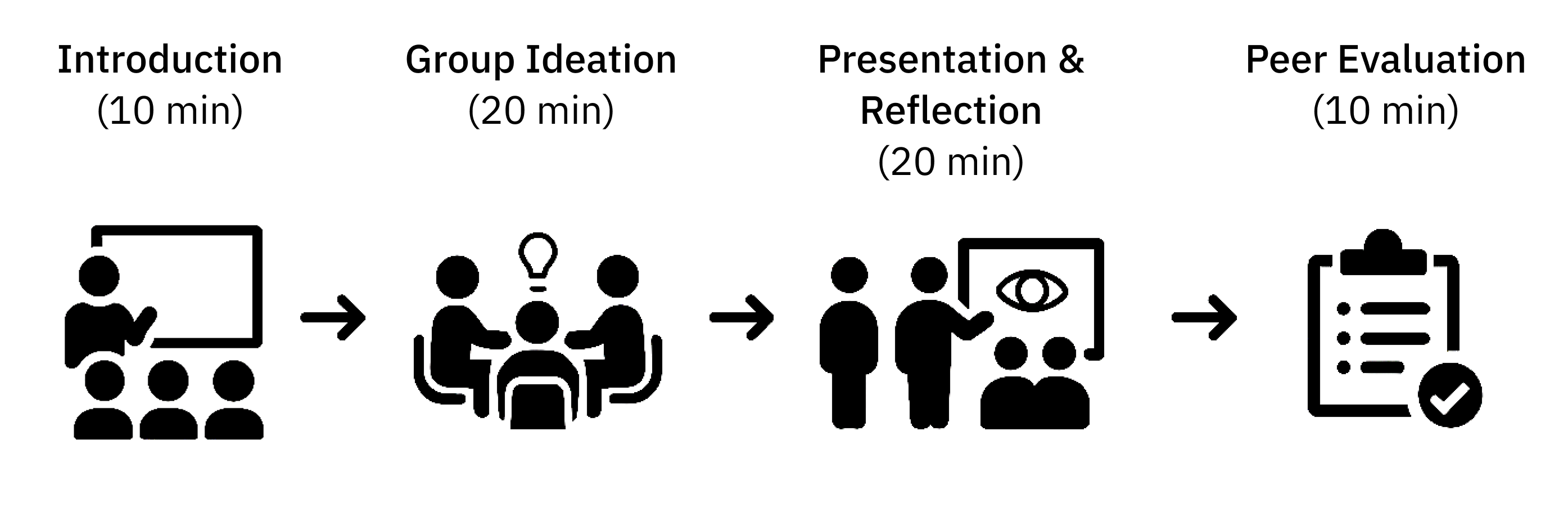}
    \caption{The structured four-step procedure used in the Phase 1 co-design workshops. Each session included a 10-minute introduction, 20 minutes for group ideation and sketching, 20 minutes for presentation and reflection, and a final 10-minute peer evaluation where groups assessed each other's designs. 
    }
    \Description{A linear timeline diagram showing the four steps of the Phase 1 workshop procedure. Step 1 is "Introduction (10 min)" represented by an icon of a person presenting to a group. Step 2 is "Group Ideation (20 min)" represented by an icon of three people at a table with a lightbulb. Step 3 is "Presentation \& Reflection (20 min)" represented by two people discussing a screen with an eye icon. Step 4 is "Peer Evaluation (10 min)" represented by a clipboard with a checkmark.}
    \label{fig:phase1_process}
\end{figure}

\textbf{Step 1: Introduction \& Task (10 min).}  
The facilitator explained study goals, ethics, and example gestures, then introduced the 9-point grid for robustness under low-precision tracking and accessibility across devices~\cite{li2017gazture, kim2022lattice, majaranta2019inducing, delamare2017designing, mollenbach2010single}. Each group received three commands and was asked to design grid-based gestures that were distinctive, memorable, and suitable for 5--7 inch mobile screens.  

\textbf{Step 2: Group Ideation (20 min).}  
Participants individually sketched gestures with rationales, then voted internally. A student assistant checked feasibility (e.g., complexity, speed, potential mimicry risks). Participants refined their ideas in a second sketching round with command labels.  

\textbf{Step 3: Presentation \& Reflection (20 min).}  
Groups presented and discussed sketches, resolving conflicts and finalizing a set of gestures. They independently rated them on a 9-point Likert scale across five dimensions: Q1 (Memorability) assessed how easily a gesture could be recalled after initial exposure~\cite{nacenta2013memorability, dong2015elicitation}; Q2 (Ease of Execution) measured the physical comfort and effort required to perform it; Q3 (Confusion / Mapping Clarity) evaluated the intuitiveness of linking the gesture's form to its intended function~\cite{sellierevaluating}; Q4 (Perceived Robustness) reflected participants' confidence in the gesture's reliability under real-world conditions~\cite{xu2018one,sherman2014user}; and Q5 (Overall Impression).  

\textbf{Step 4: Peer Evaluation Across Groups (10 min).}  
To streamline the evaluation and reduce redundancy, the four initial groups were merged into two super-groups: Group Alpha ($G_{\alpha}$) and Group Beta ($G_{\beta}$), where $G_{\alpha} = G_A \cup G_B$ and $G_{\beta} = G_C \cup G_D$. The complete gesture sets from these new super-groups were then exchanged for mutual peer evaluation ($G_{\alpha} \leftrightarrow G_{\beta}$). Each participant rated the opposing super-group's gestures on the same five dimensions, providing robust cross-validation and strengthening the reliability of the consolidated gaze gestures.

\subsection{Phase 2 Procedure} \label{method_phase2_procedure}
The second phase aimed to rigorously evaluate the refined set of user-defined gaze gestures through expert assessment, modelled after peer review systems commonly used in academic conferences. While co-design workshops with non-expert users generated a wide variety of creative gesture ideas, their feasibility, clarity, and technical robustness required expert validation. To ensure objectivity and independence, we adopted a two-step academic peer review structure:

\textbf{Independent Review.} Four experts with extensive experience in gaze interaction and eye-tracking technologies were recruited (see \S\ref{method_recruitment}) and randomly assigned to one Area Chair (AC) and three expert reviewers. Each expert independently evaluated the full set of user-generated gesture designs over a one-week review period. Review materials were provided as a compressed package, including: (1) anonymized workshop photos (with facial mosaic blurring); (2) process documentation outlining the design context and objectives; (3) a standardized evaluation template with 5-point Likert-scale and open-text fields for dimensions such as usability, novelty, and learnability; and (4) an Excel sheet containing all gesture designs formatted as labeled sticker images.

\textbf{Review Discussion Moderated by AC.} One author, serving as an additional senior expert in eye-movement and gaze interaction research, acted as AC in accordance with academic peer review conventions. While AC did not participate in the scoring, they facilitated structured discussions between reviewers. After the independent review period, a synchronous 60-minute panel discussions were held online, moderated by the AC. Reviewers were encouraged to share their evaluations, discuss disagreements, and clarify differing interpretations. Where consensus could not be reached, both majority and minority opinions were documented.

The total review process spanned approximately one week, including five days for independent assessment and two days for AC-led discussions. Final evaluations combined individual reviewer scores and panel-level insights to ensure a balanced and rigorous assessment. This multi-step expert review process allowed us to triangulate the creative outputs of non-expert participants with the critical perspectives of domain experts, thereby enhancing the robustness and practical relevance of our gesture evaluation framework.

\subsection{Data Analysis} \label{method_data_analysis}
\subsubsection{Phase 1 Data Analysis}
We video-recorded all four sessions, and the recordings were transcribed and analysed using thematic analysis~\cite{braun2006using}. All sessions were conducted in English. Three authors independently immersed themselves in the data by reading transcripts iteratively. An initial codebook was developed by jointly coding one transcript at the sentence level. A second transcript was independently coded to test the codebook, with emergent codes added through iterative discussion. The codebook was then finalised and applied to the remaining transcripts to extract themes around user motivations, design rationales, and expectations for gaze-based interaction.

In addition to qualitative data, participants collaboratively designed gaze gestures using a 9-point spatial grid and evaluated each concept via a 5-item, 9-point Likert questionnaire. These artefacts, including gesture sketches and questionnaire responses, were digitised into a structured dataset, with the following fields: participant ID (P-ID), gesture ID, intended command, and item scores for five dimensions. Proposals violating design constraints  were removed; e.g., gestures not representable on the 9-point grid, or monocular/blink-only gestures. Duplicate or convergent gestures were consolidated through name/shape matching and manual verification to yield a corpus of 102 unique gestures: 34 from Group A, 26 from Group B, 24 from Group C, and 18 from Group D (see detailed analyse in \S \ref{phase_1_overview}). These gestures and their ratings served as the foundation for Phase 2 expert evaluation. 

\subsubsection{Phase 2 Data Analysis}
We modelled the evaluation after academic peer-review conventions, leveraging the domain knowledge of four experts to refine the user-defined gestures ($n=59$) into a non-redundant, practical, curated set. Each design was indexed by a stable \texttt{PathID} (e.g., \texttt{Group $\alpha-1$}, \texttt{Group $\beta-2$}) so that expert data could be linked one-to-one with Phase 1 workshop records. We analysed two data streams: (i) numeric ratings from experts; and (ii) free-text comments ($n=237$) from the expert survey and a AC-style panel meeting (duration time=62.4 minutes). Firstly, all experts independently rated each design on three 1–5 Likert items: \emph{Usability}, \emph{Novelty}, and \emph{Learnability}. Reviewer scores were compiled and averaged per gesture across all dimensions. We computed per-design statistics (mean and standard deviation) for each dimension, as well as for a composite quality score. Inter-rater reliability was assessed using ICC, and gestures were ranked by composite mean scores, with ties resolved based on lower variance. We also analyzed 237 open-ended comments provided in the survey responses from four experts. 

In addition to numerical scores, experts provided open-ended qualitative feedback through written comments on the scoring sheets and in one AC-style review meeting. The meeting was audio-recorded and transcribed. We employed the same inductive thematic analysis approach~\cite{braun2006using} used in Phase 1 to systematically understand the experts' reasoning. Three authors first familiarized themselves with the entire corpus of feedback and transcripts. An initial codebook was then developed using a hybrid deductive–inductive approach: deductive codes were derived from established HCI and gaze interaction literature (e.g., ``ergonomic strain'', ``Midas Touch problem'', ``metaphorical consistency''). Inductive codes emerged directly from the expert data, capturing novel insights (e.g., ``critique of modality transfer'', ``context-dependent safety''). Through an iterative process of independent coding, comparison, and discussion, the codebook was refined. The final codebook consists of five core themes that represent the primary criteria used by experts to evaluate the user-generated gestures.

\subsection{Ethical Considerations} \label{method_ethics}
To address ethical concerns, we adopted multiple safeguards. First, we obtained written informed consent from all participants prior to each session. Second, we removed identifying information during analysis by pseudonymizing all participants (e.g., \texttt{S1-GA-P1} and \texttt{S2-EP1}, where \texttt{S1-GA} denotes ``Session 1--Group A'' and \texttt{S2-EP} denotes ``Session 2--Expert Participant'') and excluding any personally identifying details (e.g., real names). Third, access to the data (including interview transcripts and supporting documents such as signed consent forms and information sheets) was restricted to the authors only. All materials were stored securely in university-controlled, password-protected cloud storage. The study received ethical approval from the first author’s University Research Ethics Committee (CS16900).

\section{Phase 1: Co-Design Results} 
\begin{figure*}[!htbp]
  \centering
  \includegraphics[width=0.95\linewidth]{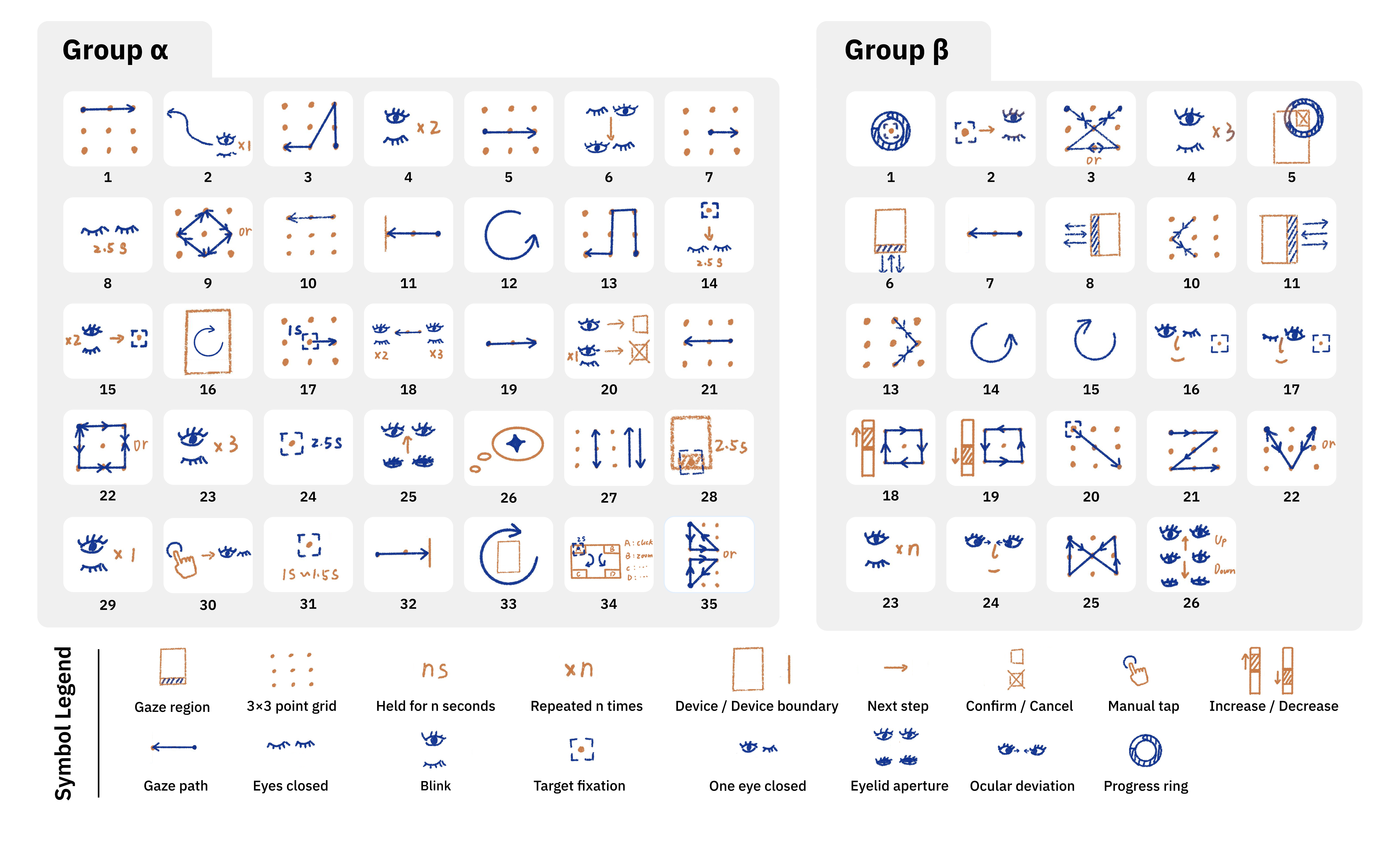}
  \caption{The complete set of 59 unique gestures generated by non-expert participants during Phase 1, shown in their standardized format. After correcting for duplicate labels (merging original gestures $\beta$-9 and $\beta$-12 into $\beta$-8 and $\beta$-11, respectively), the set was divided for peer evaluation into Group $\alpha$ (35 gestures) and Group $\beta$ (24 gestures).Full textual descriptions of each gesture, including timing and command mapping, are provided in the appendix~\ref{app:gesture_icon_definition}.}
  \Description{A large grid displaying 59 standardized icons of gaze gestures, divided into two sections: "Group Alpha" (gestures 1-35) and "Group Beta" (gestures 1-26). Each gesture is depicted using blue lines for gaze paths and orange symbols for interface elements or dwell times. A legend at the bottom explains the symbols, including: Gaze path (arrow), Eyes closed (lashes), Blink (open eye with lashes), Target fixation (brackets), One eye closed, Eyelid aperture, Ocular deviation, Progress ring, 3x3 point grid, Dwell time (ns), and Repetition (xn).}

  \label{fig:phase1_gestures}
\end{figure*}

\begin{figure*}[!htbp]
  \centering
  \includegraphics[width=0.95\linewidth]{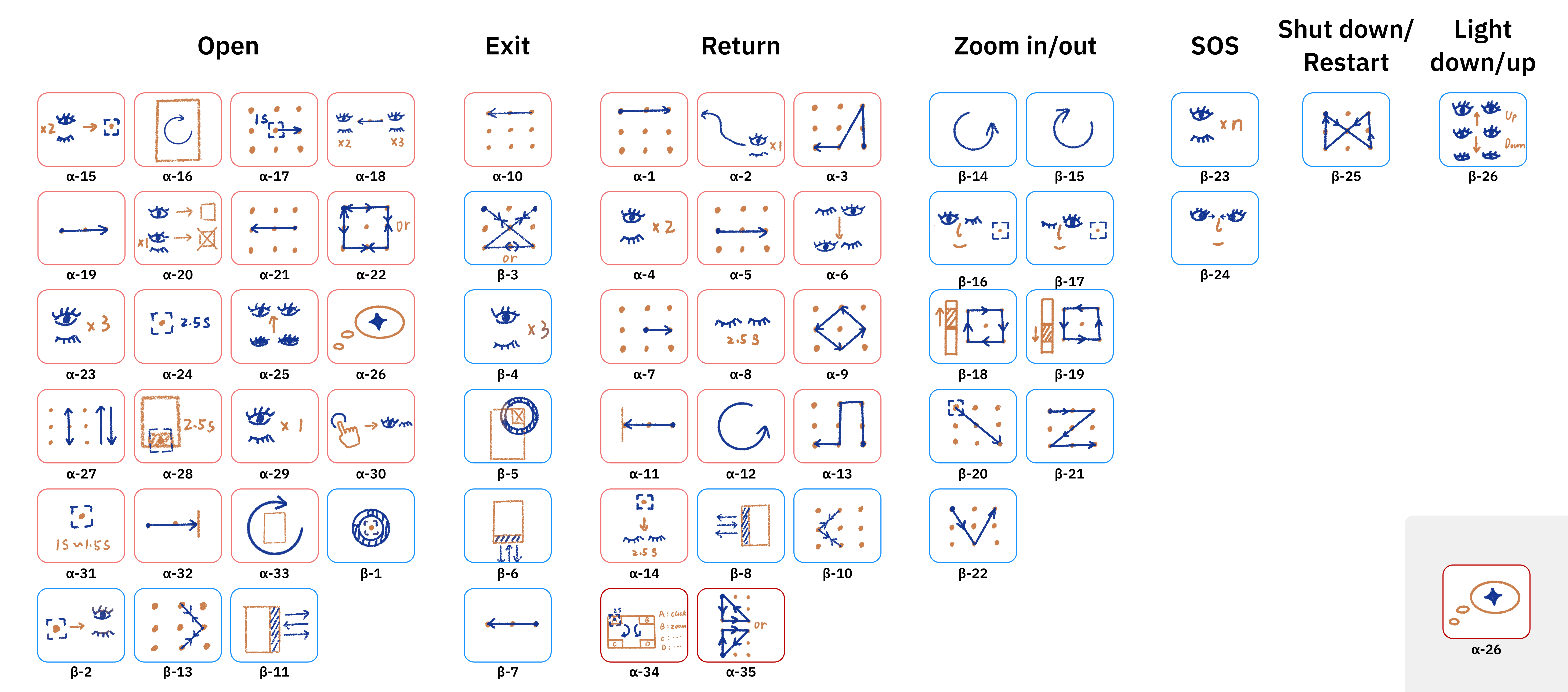}
  \caption{The user-generated gesture set from Phase 1, categorized by intended command function. This view highlights design patterns for discrete actions (e.g., Open, Exit) and continuous actions (e.g., Zoom, Light down/up). Gestures from Group $\alpha$ are denoted by red borders, and those from Group $\beta$ by blue borders. The bottom-right $\alpha$-26 gesture represents a user-defined design that does not belong to any predefined command category.}
  \Description{A chart categorizing the user-generated gestures by their intended command functions. Columns are labeled: Open, Exit, Return, Zoom in/out, SOS, Shut down/Restart, and Light down/up. Under each heading, relevant gesture icons from Group Alpha (red borders) and Group Beta (blue borders) are grouped together. For example, the "Return" column contains various leftward swipe and corner flick gestures.}
  \label{fig:phase1_by_command}
\end{figure*}

The four co-design workshops with non-expert participants from super-groups $\alpha$ and $\beta$ yielded a rich gesture set (see Fig.~\ref{fig:phase1_gestures} and Fig.~\ref{fig:phase1_by_command}). Our analysis of this creative output revealed several consistent themes: participants quickly developed sophisticated mental models for gaze control, their designs were heavily influenced by mobile device constraints and familiar metaphors, and they spontaneously invented universal modifiers to enhance usability.

\subsection{Designing Preference for Mobile Devices} 
Across all four co-design workshops, participants consistently adopted a hierarchical structure in gaze interaction design to avoid unintended commands (e.g., saving, exporting, or closing files). Rather than mapping complex actions to single continuous eye movements, participants intuitively decomposed interactions into two sequential stages: 1) target acquisition or activation, and 2) action confirmation. This approach was largely motivated by concerns over the ``Midas Touch'' problem, where unintended eye movements could trigger unwanted commands. For example, in designing a gesture for ``drag-and-drop'', a common interaction on mobile platforms, some participants proposed a sequence: dwell on the icon to select it (lift), move the gaze to the target location, then blink to confirm placement (drop). 

Nearly all participants preferred compact, edge-anchored gestures, citing ergonomic and screen space concerns. Participants gravitated toward short, efficient trajectories starting from screen borders---especially edges and corners. These designs were seen as practical and minimally intrusive. As S1-GA-P2 explained: \textit{``I already do swipes from the edge to go back on my phone. So with my eyes, it should feel the same but even quicker.''} Several echoed this, emphasizing reduced fatigue and avoiding core content areas. Gestures for commands like ``back,'' ``cancel,'' or `close'' often took the form of leftward flicks from the right bezel or diagonals from the top-left (e.g. $\alpha-1$, $\beta$-20), reinforcing familiar mobile UI mappings. Some also considered how device handling shaped feasibility. As S1-GC-P15 noted: \textit{``I'm usually holding a smaller screen in one hand. That means I'd want the gesture to happen near where I'm already looking, like the bottom or side. Not the middle, it's too tiring to move back and forth.''}

Participants then discussed interaction context, such as use while walking or multitasking. Gestures requiring minimal precision and quick activation were seen as more suitable for mobile use. Interestingly, some participants anchored gestures to physical device features---e.g., using the notch or home bar as landmarks for initiating small arcs or swipe (e.g. $\beta-22$) (S1-GB-P9). These proposals suggest that participants intuitively draw on embodied knowledge of hardware to inform gaze-based designs. During peer-groups evaluation, these compact, edge-anchored gestures (e.g.  $\alpha-1$, $\alpha-22$ and $\beta-18$) scored consistently high across all dimensions, especially ease of execution (Likert score $M=7.05, SD=1.28$), memorability (Likert score $M=7.54, SD=1.33$) and perceived robustness (Likert score $M=6.79, SD=1.48$). As S1-GD-P17 noted, they were legible, had low error potential, and aligned well with existing mobile habits: \textit{``This bottom-corner flick feels like something I'd do with a thumb already. I can see it working well for eyes too.''} However, some participants noted limitations. S1-GD-P19 observed that multi-step commands (e.g., export or save as) were harder to compress into small gestures without ambiguity. As a workaround, groups proposed using edge-based strokes as stems, modified by blinks or dwells for disambiguation (e.g. $\alpha-34$, $\beta-5$ and $\beta-11$).

\subsection{Dominance of Metaphorical Design} 
Nearly all participants consistently drew upon familiar physical or digital interaction metaphors \textbf{grounded in everyday actions} to develop their gaze-based command gestures, suggesting that users' existing motor schemas and mental models served as a natural basis for gaze control. This pattern reflects what prior gesture-elicitation work has termed a ``legacy bias''~\cite{Villarreal2020GestureElicitationReview}, in which interaction paradigms from other devices (e.g., touchscreens) act as the primary reference frame even when the input modality changes. S1-GB-P8 expressed that leveraging known patterns reduced \textit{``cognitive effort and would likely ease the learning curve for new users.''} For example, when designing a gesture for the command ``return,'' participants frequently proposed horizontal swiping motions from left to right or right to left, borrowing directly from touchscreen paradigms such as swipe-to-go-back on smartphones or trackpads (e.g. $\alpha-1$, $\alpha-32$, $\beta-7$, etc.). Rather than inventing abstract or novel gesture forms, most participants overwhelmingly preferred gestures that mirrored known physical or digital interactions, valuing usability and learnability over novelty.

\textbf{Use of symbolic and functional mapping.} In addition to physical metaphors, participants also employed \textbf{symbolic metaphors} grounded in interface design (e.g. $\alpha-12$, $\beta-10$). For the command ``copy'', several participants proposed clockwise circular motions; that is, drawing on the metaphor of rotating a dial to duplicate or produce another version. As S1-GA-P3 suggested: \textit{``Copying something feels like you're spinning it out again, like duplicating [...] I'd turn it clockwise, like a machine cranking out another version.''} In contrast, anticlockwise spirals were occasionally associated with actions like ``undo'' or ``cancel'', emphasizing the directional logic inherent in many analogue interactions, such as turning back time or reversing an operation (e.g. $\alpha-12$, $\alpha-16$). While symbolic designs were widely suggested, participants prioritized functional over purely decorative mappings. Some addressed that gaze gestures needed to communicate action rather than merely resemble visual icons. For example, S1-GD-P16 proposed an eye-shaped gesture for ``open,'' but the group rejected it, finding it lacked action-based meaning. Despite the appeal of metaphorical designs, participants noted limits in translating some metaphors into gaze interaction. Gestures familiar in touch or mouse contexts---like drag-and-drop or pinching---were often impractical without tactile feedback or fine-grained control. For instance, S1-GC-P13 and S1-GB-P7 proposed similar inward-rotating arrows to represent ``zoom out.'' (e.g. $\beta-14$) However, other participants found this gesture difficult to perform using gaze alone. As S1-GB-P9 explained: \textit{``This kind of make sense (the zoom-out gesture proposed by S1-GB-P7), I mean, it 's not as straightforward as doing a pinch gesture with my eyes, like zooming in or out in your screen. It's hard for me to imagine how that would work without using fingers.''}

\textbf{Prioritization of learnability and memorability.} Many participants justified their preference for metaphorical gestures in terms of usability goals, especially learnability and memorability. In several sessions, some participants stated that mimicking familiar gestures would reduce training time and increase trust in gaze-based systems. Furthermore, we found that during subjective evaluations, metaphor-based gestures consistently received high scores for both memorability and clarity of command mapping, suggesting strong alignment between design rationale and perceived usability. Groups tended to discard abstract or ambiguous gestures in favour of those with clear metaphorical roots. In one case, Group A eliminated a circular eye-roll gesture for ``skip'' in favour of a more direct horizontal dash gesture, as S1-GA-P1 stated: \textit{``The eye-roll feels confusing, like am I annoyed or skipping? The straight dash says it better.''}

In contrast, more creative gestures such as zigzag patterns or diagonal shapes were often rated lower, even when they appeared visually distinctive. For example, Group A proposed a spiral pattern inspired by the ``Naruto ninja forehead ribbon''. However, during the peer review of gestures, S1-GB-P5 critiqued this design, pointing out its limited usability and lack of inclusivity for older users, stating: \textit{``It's a cool shape, but I can't see myself remembering what that does. Maybe too weird [...] I don't think people like my age can finish this gesture smoothly without shaking, especially when you're walking, you need a lot of efforts.''} While metaphorical gestures dominated the workshop outcomes, there were moments of tension between innovation and familiarity. For instance, S1-GB-P8, with higher gaze interaction familiarity ($n=4$), stating: \textit{``From eye-tracking performance perspective, I know some gestures are difficult to implement reliably and hard for users to perform or interpret accurately.''} This highlights the inhibiting effect of over-reliance on metaphors, which, while safe and accessible, may constrain creativity and exploration in emerging interaction paradigms. A few participants suggested having an ``experimental'' mode in future gaze systems, where users could customize or explore abstract gestures beyond standard metaphors. Like S1-GB-P6 stated: \textit{``Why not give users an option to invent their own gestures, like advanced mode? That way you can still have defaults but also personal style.''}

\subsection{Blinks and Dwells as Universal Modifiers}
Two specific gaze-related actions---\textbf{dwelling} (maintaining gaze on a point) and \textbf{blinking}---emerged as universally proposed modifiers to enhance functionality, reliability, and intentionality. Participants saw them as essential for signalling intent, avoiding the Midas Touch problem, and structuring interaction stages (initiation, confirmation, cancellation). Their intuitive alignment with interface concepts (e.g., click, hover, escape) and potential for layering into compound commands suggest blink-dwell mechanisms should be central to practical gaze gesture systems. However, fatigue, misinterpretation, and accessibility concerns must be carefully addressed.

\textbf{Dwell: selection and initiation through sustained gaze.} Most of participants mentioned dwell as one strategy to replace ``manually tap'', defining it as holding one's gaze on a specific target or area for a predetermined duration---such as 500 to 800 ms that suggested by S1-GB-P8 and S1-GC-P12---was the most commonly suggested method for initiating or confirming interaction. Some participants also described the dwell as a natural analogue to ``hovering'' or ``pressing and holding'' with a mouse or finger. As S1-GD-P18 stated: \textit{``I think a long stare kind of means you're serious [...] you're selecting something, not just scanning around.''} While nearly all of our participants agreed that short dwell durations could serve to initiate passive actions (e.g., preview content), while longer dwell times could trigger more impactful functions, such as launching an app or confirming a command. To make dwell interactions more intuitive and less fatiguing, some participants emphasized the need for visual feedback---especially progress indicators that fill up or change colour during the dwell period. For instance, S1-GA-P3 proposed an example design: \textit{``If I'm dwelling, I'd like to see a progress bar, you know it's like a timer filling up in a game when you use an ultimate skill, so I know when it will activate.''} A few also raised the issue of involuntary dwell, especially during distracted gazes or pauses, which could result in unintended actions---a variant of the Midas Touch problem widely known in gaze interaction research.

\textbf{Blink as an explicit, click-liked confirmation mechanism.} In contrast to the passive nature of dwelling, most participants viewed blinking as a \textit{``deliberate, click-like confirmation mechanism''}---especially useful for avoiding accidental activations in gaze-only interfaces. Many participants proposed blinking as a final step to execute a command after it had been selected through gaze movement and/or dwelling. Furthermore, many participants favoured single blinks for confirmation, drawing analogies to clicking a mouse button or tapping a touchscreen. Some, like S1-GA-P4 suggested that the blink should be preceded by a dwell or a gesture to prevent unintentional clicks, or vice versa (e.g. $\alpha-15$, $\beta-2$). Several participants also proposed differentiated blink patterns to encode more complex actions. For instance, long blinks (keeping eyes closed for more than 500 ms) were proposed as a way to initiate safety-critical commands, such as locking a device or invoking an emergency protocol. As S1-GC-P11 reflected: \textit{``If I close my eyes for a bit, that's a serious signal [...] it could mean stop, or lock, something more urgent.''}  

\textbf{Complementary and challenges.} Some participants emphasized the complementary nature of blinking and dwelling. While dwelling was passive and continuous, blinking was discrete and binary. This complementarity allowed users to conceptualize interaction flows that mimicked well-established human-computer interaction patterns (e.g., hover + click, drag + drop). For instance, to open a file, S1-GB-P7 suggested dwell on the icon for 1 or 2 second (hover equivalent), and then blink once to confirm (click equivalent). S1-GD-P18 further described an effective redundancy mechanism: if a single blink with a longer dwell time serves to confirm an action, and a triple blink acts as a cancellation trigger, users gain layered control over the interaction flow, thereby reducing the likelihood of errors. Despite their utility, not all participants agreed on the reliability or comfort of blink-based interactions. A few expressed concerns about involuntary blinking, particularly in dry-eye conditions or when fatigued, which might introduce false triggers. To address this, a few participants proposed adjustable dwell durations or gaze sensitivity settings to accommodate different user needs. However, we found a few countered that blink detection algorithms in modern eye-trackers can distinguish voluntary from involuntary blinks, especially when combined with gesture context or dwell timing. S1-GB-P8 suggested that a double-confirmation mechanism---like combining dwell + blink---could mitigate such risks.  

\subsection{Innovation Recommendations}
Beyond spatial trajectories, a few participants proposed the introduction of somatic gestures---gaze commands informed by changes in eye state or facial expression. These included squinting, wide-eye opening, and even rapid alternating gaze to simulate shaking the head. They viewed these somatic inputs not only as rich interaction mechanisms but also as natural extensions of emotion-laden communication. For instance, S1-GB-P7 noted that squinting was frequently associated with reducing brightness or filtering information, while widening the eyes was proposed as a gesture to expand, zoom in, or increase visibility, stating that: \textit{``When I squint, I'm trying to reduce glare or make things smaller. It feels natural to connect that with dimming the screen or hiding details (e.g. $\beta-26$).''} Similarly, blinking intensity (soft vs. forceful) was imagined as a modifier---though a few participants acknowledged current eye-tracking technologies may not reliably capture such nuances. In particular, while innovative, these proposals sparked debates during the workshops about technical feasibility and ergonomic sustainability. Participants raised valid concerns about fatigue, especially for high-frequency actions. For example, S1-GB-P8 argued that prolonged eye opening or repetitive squinting could strain users' ocular muscles. Some also noted accessibility concerns, recognizing that not all users may be physically capable of performing certain somatic gestures due to differences in visual or facial motor function. As a result, these gestures were largely categorized as optional or supplemental interactions, rather than core commands.

Nevertheless, most participants were excited by the expressive potential of these gestures, particularly for non-traditional contexts such as gaming, immersive storytelling, or therapeutic interfaces. A few suggested the use of ``emotive eye gestures'' as part of avatar control or communication cues in virtual environments. For instance, S1-GC-P12 who had rich VR/AR device using experiences, stated that: \textit{``Imagine in a VR game mission, you squint to focus or look tough, and the character does it too. It becomes part of your identity, not just control.''} At the same time, when discussed alongside the hierarchical model, many participants proposed integrating somatic gestures as modifiers in a layered command structure. For instance, S1-GD-P17, S1-GD-P19, and S1-GD-P20 all suggested using a dwell to activate a control panel, and then a squint or wide-eye action to adjust parameters such as contrast or scale. 

\textbf{Designing for edge cases and emergencies.}
Although initial tasks focused on everyday commands (e.g., copy, paste, go back), several participants explored scenarios where gaze input could be critical, particularly in situations where other forms of input were unavailable. First, some participants proposed concept was the idea of an SOS gesture---a universal command that could be activated using only the eyes, even under duress. This took several forms, the most popular being a rapid blinking sequence, as S1-GA-P5 proposed: \textit{``If you want to make an SOS call on an iPhone, you can quickly press the side button five times. So similarly, five quick blinks could trigger a call or alert, like an SOS. Your eyes might be the only thing you can move.''} Other participants from same group (A) further discussed that these proposals were often contextualized within accessibility or health emergency scenarios, including use cases such as support for users with motor disabilities. They also proposed visual feedback for these emergency triggers, such as screen colour shifts (e.g., flashing red) or audible pings to reassure users that the signal was sent. Furthermore, several participants discussed privacy and ethical implications, suggesting that SOS gestures could be customized or disabled to avoid misuse or accidental activation. This theme linked to broader discussions on using gaze in high-stakes settings like assistive tech, surgeries, or driving. A few participants further noted that while daily tasks dominate interface design, future gaze-based systems could and should accommodate life-critical scenarios. For instance, as S1-GB-P10 stated: \textit{``I never thought of it before, but now I feel like eyes are perfect for emergencies. You don't need hands, you don't even need to move [...] just look and blink. That could save lives.''}

\begin{tapscallout}
\noindent\textbf{Findings for RQ1:} Our co-design workshop results revealed that non-expert users overwhelmingly favoured metaphorical, low-complexity gaze gestures grounded in familiar physical or digital interaction paradigms (e.g., swiping, tapping, or rotating). Participants preferred compact, edge-anchored based gestures that mirrored touchscreen behaviours and aligned with existing mental models, and valued memorability, clarity, and execution ease. A hierarchical command structure (e.g., dwell to activate, blink to confirm) emerged to reduce accidental triggers and enhance control. Symbolic and somatic elements (e.g., squinting to dim brightness) further enriched the design space, though users remained cautious about gestures requiring precision or causing fatigue. Importantly, participants also envisioned emergency and accessibility use cases, suggesting that gaze gestures should extend beyond basic commands to serve as resilient fallback mechanisms in critical contexts.
\end{tapscallout}

% \begin{tcolorbox}[colback=green!10, colframe=black!20, boxrule=0.5pt, sharp corners=southwest]
% \textbf{Findings for RQ1:} Our co-design workshop results revealed that non-expert users overwhelmingly favoured metaphorical, low-complexity gaze gestures grounded in familiar physical or digital interaction paradigms (e.g., swiping, tapping, or rotating). Participants preferred compact, edge-anchored based gestures that mirrored touchscreen behaviours and aligned with existing mental models, and valued memorability, clarity, and execution ease. A hierarchical command structure (e.g., dwell to activate, blink to confirm) emerged to reduce accidental triggers and enhance control. Symbolic and somatic elements (e.g., squinting to dim brightness) further enriched the design space, though users remained cautious about gestures requiring precision or causing fatigue. Importantly, participants also envisioned emergency and accessibility use cases, suggesting that gaze gestures should extend beyond basic commands to serve as resilient fallback mechanisms in critical contexts.
% \end{tcolorbox}

\section{Pre-Analysis of Gesture Set between Phase 1 \& Phase 2} \label{phase_1_overview}

To further examine these themes, we compared self- and peer-ratings of the gesture set, revealing alignment and divergence. Among 102 gestures, we excluded 18 gestures that violated the 9-point grid or used non-standard inputs (e.g., blinks alone), and consolidated 25 duplicates, yielding 59 unique mobile-specific gaze gestures. To support analysis, we redrawn all 59 user-designed gestures in a standardized style without altering original patterns. After correcting labelling inconsistencies (e.g., multi-step gestures split into two IDs), we obtained 35 gestures in the $\alpha$ group and 24 in the $\beta$ group (see Fig.~\ref{fig:phase1_gestures}). Gestures were further organized by intended command categories, with group membership indicated by border colour (see Fig.~\ref{fig:phase1_by_command}).

Given variability in peer-evaluation coverage where some gestures received 8 or more ratings, while others had as few as 2 or 3, we excluded any gesture with fewer than four total ratings to ensure statistical reliability. Each gesture instance was rated on 5 items (Q1--Q5). Because Q3 captures \emph{confusion} (higher indicates worse), we reverse-coded Q3 and computed a positively keyed composite $Q_{\text{pos}}$ by averaging Q1--Q5 (after reversal) so that higher values consistently indicate better perceived quality. Analysis was conducted at the \emph{gesture} level (mean across all raters per gesture) and contrasted \emph{self-ratings} against \emph{peer-ratings} using paired tests. We first averaged ratings \emph{within gesture} to obtain per-gesture means and then compared \emph{across gestures} with paired tests (one paired row per gesture). We then used paired $t$-tests to test item-wise differences on per-gesture means. For a directionally consistent composite, Q3 was reverse-coded ($Q3' = 10 - Q3$) and a \emph{positively keyed} composite $Q_{\text{pos}}$ was computed as the mean of Q1--Q5 (with $Q3'$). For the composite we report paired $t$, $p$, Cohen's $d_{z}=t/\sqrt{N}$, and 95\% CIs of the mean difference. Self--peer \emph{agreement} in gesture ordering was summarised by Pearson's $r$ across gestures. Finally, we list gestures with the largest absolute self--peer gaps on Q5 to prioritise redesign. Normality of difference scores was inspected; results were robust to non-parametric checks.

We found that Group $\alpha$'s self and peer scores are broadly aligned (see Table \ref{tab:itemwise_combined}). Participants reported slightly less confusion (Q3) and higher robustness (Q4) than peers, while memorability, execution ease, and overall impression were statistically indistinguishable. In contrast, Group $\beta$ shows a strong \textit{self-enhancement} pattern: self-ratings exceed peer ratings by approximately 1 point on Q1, Q2, Q4, and Q5. Notably, regarding Q3 (pre-reversal), participants in Group $\beta$ also perceived significantly greater confusion ($\Delta \approx 2.28$) compared to peers. Regarding composite scores, Table \ref{tab:stats_phase1_combined} illustrates that Group A shows a small, non-significant self-advantage, whereas Group B exhibits a clear, large self-advantage ($d_z=1.43$) with narrow confidence intervals. Furthermore, the table reveals that self and peer ratings only weakly agree on the relative ordering of gestures ($r \le 0.31$), supporting the need to report both item-level and composite outcomes and to review outliers qualitatively. Finally, Table \ref{tab:disagreements_phase1_all} shows gestures exhibiting the strongest divergences between self and peer on overall quality; these are high-value candidates for clarification (naming/mapping), simplification, or additional instruction before subsequent testing.

\begin{table}[ht]
\centering
% \fontsize{7pt}{8pt}\selectfont
\setlength{\tabcolsep}{2pt}
\resizebox{0.99\columnwidth}{!}{
\begin{tabular}{l ccc c ccc}
\toprule
& \multicolumn{3}{c}{\textbf{Group $\alpha$ (GA \& GB)}} & \phantom{a} & \multicolumn{3}{c}{\textbf{Group $\beta$ (GC \& GD)}} \\
\cmidrule{2-4} \cmidrule{6-8}
\textbf{Item} & \textbf{Self / Peer} & \textbf{$\Delta$} & \textbf{$t$ ($p$)} & & \textbf{Self / Peer} & \textbf{$\Delta$} & \textbf{$t$ ($p$)} \\
\midrule
Q1 & 7.50 / 7.35 & +0.16 & 0.93 (.360) & & 8.15 / 7.12 & +1.04 & 8.80 ($<$.001) \\
Q2 & 7.18 / 7.35 & -0.17 & -0.92 (.362) & & 7.62 / 6.57 & +1.05 & 6.18 ($<$.001) \\
Q3 & 5.51 / 6.03 & -0.52 & -2.57 (.015) & & 7.79 / 5.51 & +2.28 & 13.41 ($<$.001) \\
Q4 & 6.94 / 6.53 & +0.41 & 2.21 (.034) & & 7.67 / 6.54 & +1.13 & 9.55 ($<$.001) \\
Q5 & 6.43 / 6.41 & +0.01 & 0.09 (.931) & & 7.60 / 6.29 & +1.31 & 9.96 ($<$.001) \\
\bottomrule
\end{tabular}
}
\caption{Item-wise paired comparison (Self vs. Peer) for Group $\alpha$ ($N{=}34$) and Group $\beta$ ($N{=}26$). $\Delta=\text{Self}-\text{Peer}$.}
\label{tab:itemwise_combined}
\end{table}

\begin{table*}[ht]
\centering
% \fontsize{7pt}{8pt}\selectfont
\setlength{\tabcolsep}{3pt} % Adjust padding to fit width
\resizebox{0.95\linewidth}{!}{
\begin{tabular}{l c ccc l l c cc}
\toprule
& & \multicolumn{5}{c}{\textbf{Composite Score Differences ($Q_{\text{pos}}$)}} & \phantom{a} & \multicolumn{2}{c}{\textbf{Self-Peer Agreement ($r$)}} \\
\cmidrule{3-7} \cmidrule{9-10}
\textbf{Target Set} & \textbf{$N$} & \textbf{Self} & \textbf{Peer} & \textbf{$\Delta$} & \textbf{$t$ ($p$)} & \textbf{$d_{z}$ (95\% CI of $\Delta$)} & & \textbf{on Q5} & \textbf{on $Q_{\text{pos}}$} \\
\midrule
Group $\alpha$ (GA \& GB) & 34 & 6.51 & 6.32 & +0.19 & 1.72 ($p{=}0.094$) & 0.30 $[-0.034,\; 0.405]$ & & 0.038 & 0.250 \\
Group $\beta$ (GC \& GD) & 26 & 6.65 & 6.20 & +0.45 & 7.28 ($p{<}0.001$) & 1.43 $[0.321,\; 0.575]$ & & 0.308 & 0.092 \\
\bottomrule
\end{tabular}
}
\caption{Composite score analysis ($Q_{\text{pos}}$, Q3 reversed) including effect sizes, paired with Self-Peer Agreement correlations across gestures (Pearson $r$).}
\label{tab:stats_phase1_combined}
\end{table*}

% \medskip
% % -------------------------
% % TABLE 3: COMPOSITE
% % -------------------------
% \begin{table}[t]
% \centering
% \fontsize{8pt}{8pt}\selectfont
% \begin{tabular}{lrccrll}
% \toprule
% \textbf{Target set} & \textbf{$N$} & \textbf{Self} & \textbf{Peer} & \textbf{$\Delta$} & \textbf{$t$ ($p$)} & \textbf{$d_{z}$ (95\% CI of $\Delta$)} \\
% \midrule
% GA \& GB gestures & 34 & 6.508 & 6.323 & +0.185 & 1.72 ($p{=}0.094$) & 0.30 \; ($[-0.034,\;0.405]$) \\
% GC \& GD gestures & 26 & 6.648 & 6.200 & +0.448 & 7.28 ($p{\approx}1.26{\times}10^{-7}$) & 1.43 \; ($[0.321,\;0.575]$) \\
% \bottomrule
% \end{tabular}
% \caption{Paired tests on the positively keyed composite ($Q_{\text{pos}}$; Q3 reversed).}
% \label{tab:composite_phase1_all}
% \end{table}

% \medskip
% % -------------------------
% % TABLE 4: AGREEMENT
% % -------------------------
% \begin{table}[t]
% \centering
% \fontsize{8pt}{8pt}\selectfont
% \begin{tabular}{lcc}
% \toprule
% \textbf{Target set} & \textbf{$r$ on Q5} & \textbf{$r$ on $Q_{\text{pos}}$} \\
% \midrule
% GA \& GB gestures & 0.038 & 0.250 \\
% GC \& GD gestures & 0.308 & 0.092 \\
% \bottomrule
% \end{tabular}
% \caption{Self--peer agreement across gestures (Pearson $r$).}
% \label{tab:agreement_phase1_all}
% \end{table}

% \medskip
% -------------------------
% TABLE 5: TOP DISAGREEMENTS
% -------------------------
\begin{table}[ht]
\centering
% \fontsize{7pt}{8pt}\selectfont
\resizebox{0.99\columnwidth}{!}{
\begin{tabular}{lrccr}
\toprule
\textbf{Target set} & \textbf{Gesture} & \textbf{Self} & \textbf{Peer} & \textbf{$\Delta$} \\
\midrule
GA \& GB gestures & \#27 & 6.57 & 4.50 & +2.07 \\
      & \#7  & 7.29 & 5.33 & +1.95 \\
      & \#18 & 6.14 & 4.83 & +1.31 \\
      & \#22 & 7.14 & 5.83 & +1.31 \\
      & \#4  & 5.57 & 6.83 & $-$1.26 \\
\midrule
GC \& GD gestures & \#5  & 8.71 & 5.86 & +2.86 \\
      & \#24 & 7.00 & 4.57 & +2.43 \\
      & \#1  & 9.00 & 7.00 & +2.00 \\
      & \#8  & 8.14 & 6.14 & +2.00 \\
      & \#7  & 7.57 & 5.71 & +1.86 \\
\bottomrule
\end{tabular}
}
\caption{Largest absolute self--peer gaps on Q5 (per gesture). Positive $\Delta$ indicates Self$>$Peer.}
\label{tab:disagreements_phase1_all}
\end{table}

\begin{tapscallout}
\noindent\textbf{Findings for RQ1:} Our analysis of the 59 user-defined gaze gestures revealed that non-expert participants predominantly designed gestures that were metaphorically grounded, visually distinct, and easy to perform. Self-ratings were generally higher than peer-ratings across most dimensions, particularly ease of execution, robustness, and overall impression. Interestingly, participants also reported higher confusion scores (pre-reversal), indicating a degree of critical self-awareness. While one design set showed a significant self-enhancement in composite scores, the other showed only a modest, non-significant difference. Overall, alignment between self and peer ratings was low, with weak agreement in gesture rankings. Gestures with large self-peer discrepancies, especially in overall impression, were flagged as candidates for redesign.
\end{tapscallout}

% \begin{tcolorbox}[colback=green!10, colframe=black!20, boxrule=0.5pt, sharp corners=southwest]
% \textbf{Findings for RQ1:} Our analysis of the 59 user-defined gaze gestures revealed that non-expert participants predominantly designed gestures that were metaphorically grounded, visually distinct, and easy to perform. Self-ratings were generally higher than peer-ratings across most dimensions, particularly ease of execution, robustness, and overall impression. Interestingly, participants also reported higher confusion scores (pre-reversal), indicating a degree of critical self-awareness. While one design set showed a significant self-enhancement in composite scores, the other showed only a modest, non-significant difference. Overall, alignment between self and peer ratings was low, with weak agreement in gesture rankings. Gestures with large self-peer discrepancies, especially in overall impression, were flagged as candidates for redesign.
% \end{tcolorbox}

% ----------------------------------------------------------------------------------------------------------------------------------------------------------------------------------------------------
% Phase 2
% ----------------------------------------------------------------------------------------------------------------------------------------------------------------------------------------------------

\section{Phase 2: Expert Peer-Review Evaluation} \label{result_phase_2}
\subsection{Qualitative Results from Expert Evaluation}
\label{sec:phase2_qual}
\begin{figure*}[!htbp]
  \centering
  \includegraphics[width=0.95\linewidth]{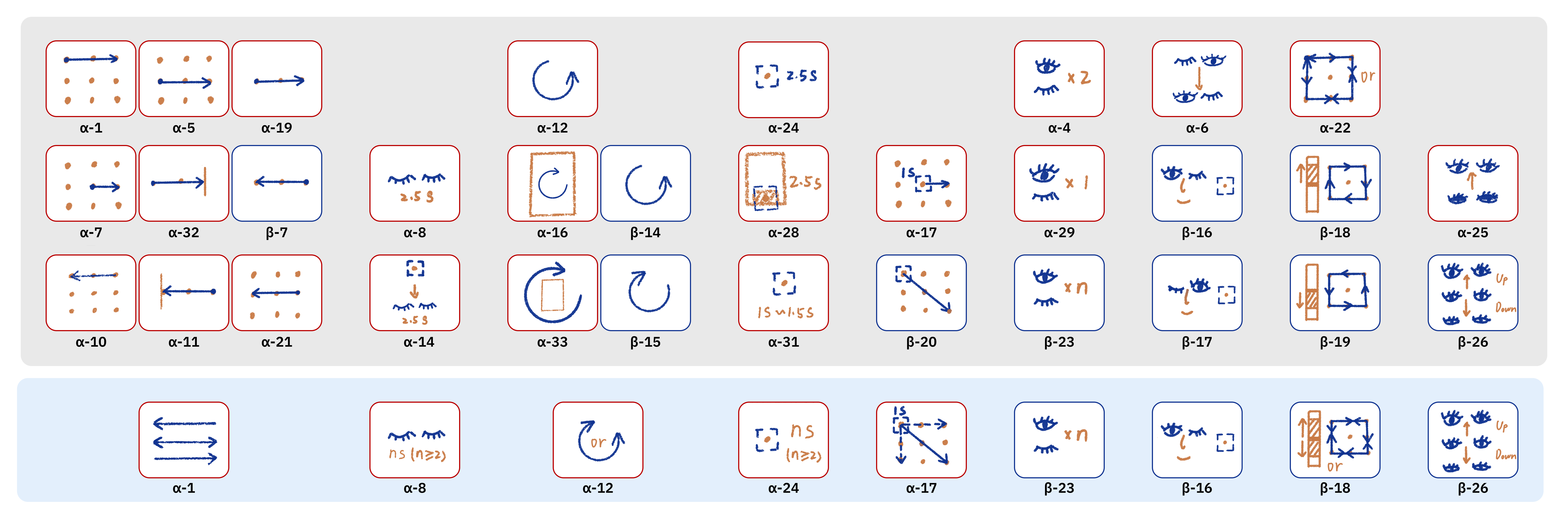}
  \caption{The gesture consolidation process during the expert review discussion. Functionally and visually similar designs, such as these horizontal swipes, were merged into a single representative gesture to create a non-redundant set.
  }
  \Description{An illustration of the gesture consolidation process. The top row shows three variations of horizontal swipe gestures (short strokes, dot-grid paths, and long strokes). An arrow points downward to a single, unified icon representing a standardized horizontal left-to-right swipe, demonstrating how multiple similar user sketches were merged into one representative gesture.}
  \label{fig:merge_example}
\end{figure*}

\begin{figure*}[ht]
  \centering
  \includegraphics[width=0.85\linewidth]{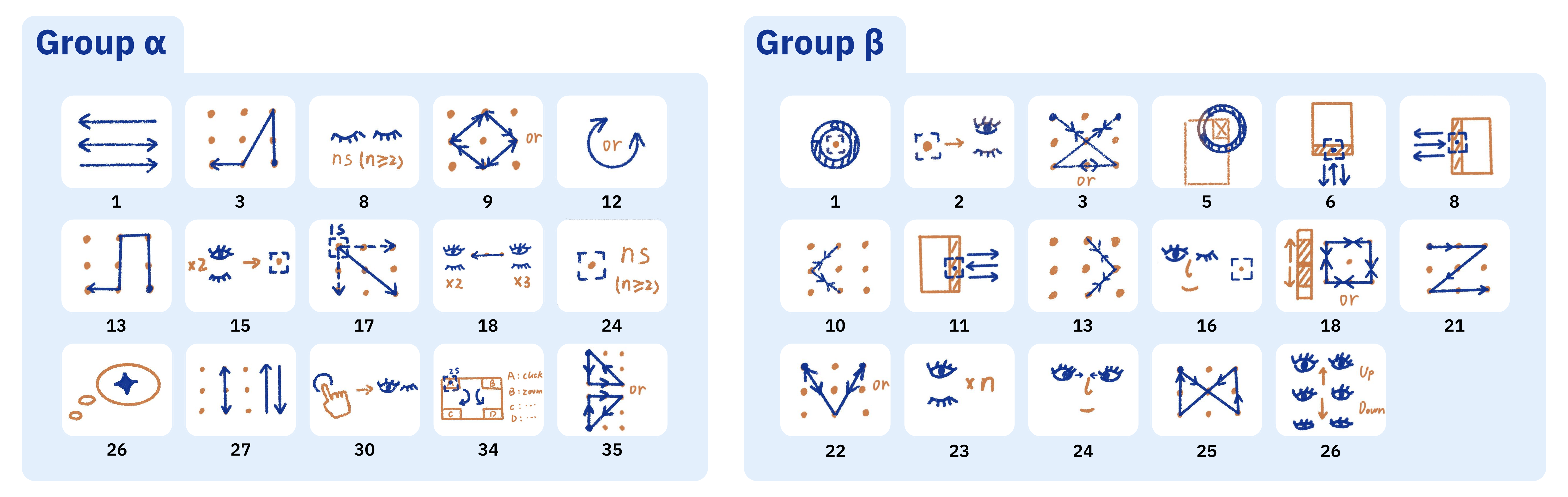}
  \caption{The final expert-validated gesture set, containing 32 robust designs refined from the initial user-generated concepts. This set is composed of 15 gestures from Group $\alpha$ and 17 from Group $\beta$, representing designs that experts deemed ergonomically sound, semantically clear, and technically viable.
  }
  \Description{A grid displaying the final set of 32 expert-validated gaze gestures. The set is divided into "Group Alpha" (15 gestures) and "Group Beta" (17 gestures). The icons depict various eye movements including cardinal swipes, corner dwells, geometric shapes (circles, triangles), and blink sequences, drawn in the standardized blue and orange format.}
  \label{fig:final_gestures}
\end{figure*}

To ensure a robust and non-redundant curated gaze gestures, we consolidated highly similar gestures into unified representative forms. For example, as shown in Fig.~\ref{fig:merge_example}, leftward and rightward movements of varying lengths (top) were merged into a single standardized pattern (bottom). After the expert peer-review process, a total of 32 gestures were retained, including 15 from the $\alpha$ group and 17 from the $\beta$ group (see Fig.~\ref{fig:final_gestures}). 
From this process, three primary objectives emerged that characterize a successful gaze gesture from an expert perspective: (1) Human Factors, (2) Interaction Design, and (3) Technical \& Contextual Factors.

\subsubsection{Human Factors: Prioritizing Ergonomic and Cognitive Plausibility}
All experts highlighted the fundamental human constraints of the ocular system. While users in Phase 1 proposed a wide variety of visually creative shapes, experts consistently rejected those that were not ergonomically viable, prioritizing physiological comfort over aesthetic novelty. Firstly, they emphasized that ocular motor control is primarily saccadic. They noted the eye \textit{``saccade movement is a kind of ballistic motion''} and is ill-suited for tracing curves without a visual guide (\textbf{A1.1 Saccadic Affinity}, \textbf{A1.2 Smooth Pursuit Limitation}). Consequently, gestures involving spirals or free-form arcs (e.g., \texttt{$\alpha-2$} and \texttt{$\beta-14$} in Fig~\ref{fig:phase1_gestures}) were consistently criticized as imprecise and likely to cause \textbf{Ocular Muscular Strain \& Fatigue (A1.3)}. Even for linear paths, experts favoured cardinal directions (up/down, left/right) over diagonals (e.g., in \texttt{$\beta-22$}), as the latter require more complex ocular muscle coordination, involving four or more muscles instead of just two. As EP3 stressed: \textit{``Eye paths are not like a drawing, it's not designed to smoothly trace curves or diagonals for long [...] fatigue sets in very quickly.''} 

Second, all experts highlighted gestures that imposed a high cognitive load and demonstrated issues with internal consistency. In particular, three experts independently raised concerns about the problem of \textbf{Input/Output Conflict, also known as the Midas touch problem (A2.2)}, where a user must simultaneously execute a precise gesture (output) while visually monitoring its effect (input). This dual demand places excessive strain on users' perceptual and motor resources, reducing both accuracy and usability. For example, EP2 described this as a major flaw in continuous zooming gestures: \textit{``While zooming, the eye needs to see the content as well, the Midas touch problem will be happened... this action might exceed the human eye's range of capability.''} This insight underscores the importance of designing gaze gestures that decouple control actions from simultaneous perceptual monitoring, thereby minimizing cognitive overload and avoiding conflicts between input and feedback channels.

\begin{tapsbox}[gray!10]
\noindent\textbf{Design Heuristic 1:} Favor gestures composed of short, cardinal (horizontal/vertical) strokes. Avoid long curves, circles, and repeated zig-zags, especially for frequent commands.

\noindent\textbf{Design Heuristic 2:} Decouple complex commands from continuous visual inspection. Use a discrete action like a dwell or blink to switch between an ``inspection'' mode and a ``commanding'' mode. Alternatively, employ specific eye movements such as rapid up-down/left-right scanning, involving only muscular motion while disregarding information input.
\end{tapsbox}

% \begin{tcolorbox}[colback=gray!10, colframe=black!20, boxrule=0.5pt, sharp corners=southwest]
% \textbf{Design Heuristic 1:} Favor gestures composed of short, cardinal (horizontal/vertical) strokes. Avoid long curves, circles, and repeated zig-zags, especially for frequent commands.

% \textbf{Design Heuristic 2:} Decouple complex commands from continuous visual inspection. Use a discrete action like a dwell or blink to switch between an ``inspection'' mode and a ``commanding'' mode. Alternatively, employ specific eye movements such as rapid up-down/left-right scanning, involving only muscular motion while disregarding information input.
% \end{tcolorbox}

\subsubsection{Interaction Design: Balancing Metaphor, Ambiguity, and Structure}
Experts evaluated the designs based on established principles of interaction design, validating strong user intuitions while critiquing flawed assumptions.

\textbf{Semantics \& metaphor.} All experts noted that metaphors need to strike a balance between semantic resonance and technical feasibility. They endorsed gestures grounded in universal UI and bodily metaphors (\textbf{Direct Metaphor Transfer, C2.1}) consistently scored well, particularly when they drew on widely recognized symbolic conventions. For example, using a `X' shape to close (\texttt{$\beta-3$}) or a horizontal line to delete mapped directly onto established touchscreen or desktop interactions (\texttt{$\alpha-11$, $\beta-7$}). Experts emphasized that such direct mappings reduced the learning curve, as users could easily carry over pre-existing knowledge from other interfaces. However, experts were equally critical when metaphors failed to transfer across modalities. Three experts specifically raised the issue of \textbf{Modality Mismatch Critique (C2.3)}, exemplified by a gesture that attempted to mimic the two-finger `pinch-to-zoom' interaction. While intuitive on touchscreens, experts argued it was impractical for gaze-based input, since gaze lacks two independently controllable points of reference. In contrast, \textbf{Somatic Metaphors (C2.2)}---gestures inspired by bodily expressions---were seen as highly innovative. All experts praised the gesture of squinting to dim the screen (\texttt{$\beta-26$}) for its semantic clarity, as squinting naturally corresponds to the concept of ``making things darker.'' Yet, experts also concerned that such metaphors would require robust calibration, since natural squints often occur spontaneously under bright light or fatigue. As EP2 noted: \textit{``For developers, this is difficult to recognize [...] I mean, the system needs to ensure it doesn't confuse real squinting with accidental squinting. Otherwise, you risk frustrating users with false triggers.''} 

\textbf{Distinctiveness \& ambiguity.} All experts stressed the need for gestures to exhibit a \emph{distinctive signature} that could be reliably distinguished from the ``noise'' of everyday visual behaviour (\textbf{Signal Uniqueness vs. Noise, B1.1}). The eye naturally produces micro-saccades, smooth pursuits, and fixations that resemble intentional gestures. For this reason, gestures that were overly simple---such as single horizontal sweeps aligned with reading direction (e.g., \texttt{$\alpha-1$})---were flagged as high-risk for false positives.  To address this, experts consistently recommended \textbf{Disambiguation Strategies (C1.2)}. For instance, instead of a single sweep, a deliberate left-right-left zigzag would be much harder to confuse with natural scanning (e.g. $\alpha-1$ in Fig~\ref{fig:merge_example}). Another proposed disambiguation technique was to combine directionality with dwell or blink markers, effectively layering multiple channels of intent. Furthermore, EP2 and EP4 all noted the need to keep gestures simple enough to avoid fatigue while still distinctive enough to ensure reliability: \textit{``Users don't want to draw complicated symbols with eyes, but also can't rely on single strokes. So this isn't just a technical issue; it's also a cognitive and ergonomic one. Designers must carefully calibrate between naturalistic movements and artificial signals, ensuring gestures are both memorable and reliably machine-detectable.''} [EP4]

\textbf{Interaction structure (compositionality).} All experts acknowledged the effectiveness of two-stage dwell-to-arm → act-to-commit grammar, recognizing it as a fundamental prerequisite for reliability, particularly for mitigating the Midas Touch (unintended activations). For example, designs that followed this rule---such as, \texttt{$\alpha-30$, $\beta-1$, $\beta-2$}---were rated highest for usability and feasibility. Experts praised these gestures' ability to restore a sense of agency. And, the $\alpha-24$, $\alpha-28$ and $\alpha-31$ in Fig~\ref{fig:phase1_gestures} reveal participants' confusion regarding dwell duration. Excessively brief pauses make it difficult to distinguish between viewing and command execution, the Midas Touch confusion; thus, dwell duration should ideally exceed 2 seconds, i.e. the $\alpha-24$ in Fig~\ref{fig:merge_example}. Beyond mitigating accidental triggers, all experts argued that compositionality could serve as a foundation for more complex hierarchical command structures. For instance, EP4 added: \textit{``A simple gesture is like a basic vocabulary, it can give users a high degree of customization [...] You can chain gestures like dwell, then direction, then blink. Then the user can build a whole grammar.''} However, EP1 and EP2 argued that compositionality must remain lightweight, noting that overly elaborate multi-stage gestures risk slowing interaction and increasing error rates. 

\begin{tapsbox}[gray!10]
\noindent\textbf{Design Heuristic 3:} To prevent accidental activation, ensure simple gestures have a distinctive signature from the natural eye movements, either through repetition (e.g., L-R-L), angular paths (e.g., a chevron) and specific designed paths (e.g., timer), or by combining them with another primitive.
\end{tapsbox}

% \begin{tcolorbox}[colback=gray!10, colframe=black!20, boxrule=0.5pt, sharp corners=southwest]
% \textbf{Design Heuristic 3:} To prevent accidental activation, ensure simple gestures have a distinctive signature from the natural eye movements, either through repetition (e.g., L-R-L), angular paths (e.g., a chevron) and specific designed paths (e.g., timer), or by combining them with another primitive.
% \end{tcolorbox}

\subsubsection{Technical \& Contextual Factors: Implementation and Adaptation}
Experts evaluated the user-defined gaze gestures not only through the lens of physiological plausibility and interaction design, but also by intuitively assessing their real-world feasibility for system implementation.

\textbf{System Implementation \& Calibration.} All experts emphasized several key implementation requirements that often went overlooked in user-generated designs. For example, blink-based commands---such as those in \texttt{$\beta-4$} and \texttt{$\beta-23$}---were praised for their intuitive semantics (e.g., blink to select or confirm), but flagged for their vulnerability to misinterpretation without strict timing constraints. This challenge was labelled as \textbf{Inter-Gesture Confusability (B1.2)}. Similarly, somatic gestures like squinting or eye widening were viewed as promising for their metaphorical clarity but raised concerns about \textbf{Personalization and Calibration (B2.3)}. Experts pointed out that the default eye aperture varies significantly across users due to facial anatomy, age, and even lighting conditions. For instance, EP4 stressed the necessity of building in per-user calibration phases, similar to those used in biometric systems or eye-tracking experiments, stating: \textit{``For someone with smaller eyes, their normal might already look like a squint. If the system isn't calibrated for each person, the gesture either won't trigger at all or will trigger constantly (e.g. $\alpha-25$ and $\beta-26$).''} The $\alpha-8$ and $\alpha-14$ in Fig~\ref{fig:phase1_gestures} exhibit similar cases. The $\alpha-14$ gesture consists of a fixation at a specific location, followed by closing the eyes for 2.5 seconds, which makes it difficult for the system to distinguish the precise point of gaze prior to closure. Conversely, $\alpha-8$ merely detects the 2.5-second eye closure, which is easier to identify. In addition, all experts argued the continuous gestures, such as sliders or zooming actions exemplified in \texttt{$\beta-18$}, invoking the constraint of \textbf{Modality Precision Limits (B2.2)}, noting that while eyes are excellent for saccadic pointing, they are poorly suited for precise, smooth tracking over time, as saccadic process are easily interrupted by external stimuli. Experts thus recommended that such fine-grained tasks be offloaded to other modalities (e.g., voice, head movement, or touch) when possible.

\textbf{Context-Sensitive Complexity.} All experts stressed the importance of \textbf{Proportional Complexity (C3.1)}---the idea that gesture complexity should scale with the importance or risk of the action it triggers. While user participants tended to value simplicity and speed, experts challenged the assumption that \textit{``simpler is always better.''} Instead, they proposed a design rule where high-stakes, low-frequency actions should require more deliberate and effortful gestures to prevent accidental activation. For example, \texttt{$\beta-25$} introduced a multi-part gesture to trigger a system shutdown. Although some users viewed it as cumbersome, EP3 defended its value, stating: \textit{``The complexity of this pattern corresponds to the sensitivity of the operation, no one want to shut down the system with a flick.''} EP2 also linked the interaction cost with operational risk, noted: \textit{``For shutting down, or even logging out, user need a bit complicated operations, this is not a bad design, it's a good risk management.''} Experts further noted that complexity can also act as a confirmatory mechanism in multi-user or high-distraction environments, such as smart home household contexts.

\begin{tapsbox}[gray!10]
\noindent\textbf{Design Heuristic 4:} Provide progress feedback for dwells, implement personal calibration for somatic gestures, and use multi-modal input (e.g., gaze + voice, gaze + touch, gaze + IMU, etc.) for fine-grained continuous control.
\end{tapsbox}

% \begin{tcolorbox}[colback=gray!10, colframe=black!20, boxrule=0.5pt, sharp corners=southwest]
% \textbf{Design Heuristic 4:} Provide progress feedback for dwells, implement personal calibration for somatic gestures, and use multi-modal input (e.g., gaze + voice, gaze + touch, gaze + IMU, etc.) for fine-grained continuous control.
% \end{tcolorbox}

\subsection{Quantitative Results from Expert Evaluation}
\label{results_phase2_scales}

Our quantitative results showed low agreement for \emph{Usability} and \emph{Learnability}, and moderate agreement for \emph{Novelty} (see Table~\ref{tab:icc_phase2_link}). Internal consistency was acceptable (Cronbach’s~$\alpha$ reported separately). Analysis of per-rater means and SDs (Table~\ref{tab:rater_leniency}) revealed systematic leniency/severity effects, motivating the use of per-design averages and reporting of dispersion metrics. Expert rankings were based on the composite means; ties were resolved by lower SD (higher agreement) and then by greater rater count. Table~\ref{tab:expert_ranking} present the top- and bottom-ranked gestures respectively, alongside dispersion and coverage. To compare expert and participant assessments, we computed a workshop composite score per gesture using Q1–Q5 from the Phase~1 questionnaire (on 1–9 scales), reverse-coding Q3 (\emph{Confusion}) and averaging across raters (self and peer). Because the rating scales differed across phases, all alignment analyses were conducted on ranks. We also found that rank correlation was weak: Spearman’s $\rho = -0.085$; Kendall’s $\tau = -0.066$, indicating that experts and participants often prioritised different design qualities. We further computed $\Delta_{\text{rank}} = \mathrm{Rank}{\text{workshop}} - \mathrm{Rank}{\text{expert}}$ for gesture-level comparison, with negative values indicating higher participant preference. Table~\ref{tab:rank_discrepancies} enumerates gestures with the largest discrepancies.

\begin{table}[ht]
  \centering
  % \fontsize{7pt}{8pt}\selectfont
  % \resizebox{0.8\columnwidth}{!}{
  \begin{tabular}{lcccc}
    \toprule
    \textbf{Dimension} & \textbf{ICC(2,1)} & \textbf{ICC(2,$k$)} & \textbf{$N$} & \textbf{$k$} \\
    \midrule
    Usability     & 0.042 & 0.117 & 50 & 3 \\
    Novelty       & 0.300 & 0.562 & 50 & 3 \\
    Learnability  & 0.088 & 0.225 & 50 & 3 \\
    \bottomrule
  \end{tabular}
  % }
  \caption{Inter-rater reliability among experts.}
  \label{tab:icc_phase2_link}
\end{table}

\begin{table}[ht]
  \centering
  \fontsize{7pt}{8pt}\selectfont
  \resizebox{0.6\columnwidth}{!}{
  \begin{tabular}{lccc}
    \toprule
    \textbf{Rater} & \textbf{Count} & \textbf{Mean} & \textbf{SD} \\
    \midrule
    S2-EP1 & 59 & 3.73 & 0.99 \\
    S2-EP2 & 59 & 3.56 & 0.82 \\
    S2-EP3 & 59 & 2.64 & 0.85 \\
    \bottomrule
  \end{tabular}
  }
  \caption{Rater consistency summary (leniency/scale use).}
  \label{tab:rater_leniency}
\end{table}

% \begin{table*}[h]
%     \centering
%     \fontsize{7pt}{8pt}\selectfont
%     \begin{minipage}{.55\linewidth}
%       \centering
%         \begin{tabular}{lcccc}
%             \toprule
%             \textbf{Dimension} & \textbf{ICC(2,1)} & \textbf{ICC(2,$k$)} & \textbf{$N$} & \textbf{$k$} \\
%             \midrule
%             Usability & 0.042 & 0.117 & 50 & 3 \\
%             Novelty & 0.300 & 0.562 & 50 & 3 \\
%             Learnability & 0.088 & 0.225 & 50 & 3 \\
%             \bottomrule
%         \end{tabular}
%         \caption{Inter-rater reliability among experts}
%         \label{tab:icc_phase2_link}
%     \end{minipage}%
%     \hfill
%     \begin{minipage}{.40\linewidth}
%       \centering
%         \begin{tabular}{lccc}
%             \toprule
%             \textbf{Rater} & \textbf{Count} & \textbf{Mean} & \textbf{SD} \\
%             \midrule
%             S2-EP1 & 59 & 3.73 & 0.99 \\
%             S2-EP2 & 59 & 3.56 & 0.82 \\
%             S2-EP3 & 59 & 2.64 & 0.85 \\
%             \bottomrule
%         \end{tabular}
%         \caption{Rater consistency summary}
%         \label{tab:rater_leniency}
%     \end{minipage} 
% \end{table*}

\begin{table}[ht]
\centering
% \fontsize{7pt}{8pt}\selectfont
\begin{tabular}{lccllcc} 
\toprule
\multicolumn{3}{c}{Top 10 designs}   &  & \multicolumn{3}{c}{Bottom 10 designs}            \\ 
\cline{1-3}\cline{5-7}
\textbf{PathID} & \textbf{Mean} & \textbf{SD} &  & \textbf{PathID} & \textbf{Mean} & \textbf{SD}  \\ 
\cmidrule{1-3}\cmidrule{5-7}
$\beta$-23   & 4.78  & 0.38  &  &  $\beta$-21 & 1.83 & 0.71    \\
$\beta$-4    & 4.67  & 0.33  &  &  $\alpha$-26 & 2.00 & 1.41      \\
$\alpha$-15  & 4.67  & 0.47  &  &  $\alpha$-7  & 2.22 & 0.19      \\
$\beta$-12   & 4.33  & 0.49  &  &  $\alpha$-19 & 2.22 & 0.19    \\
$\beta$-26   & 4.11  & 1.26  &  &  $\alpha$-29 & 2.33 & 0.33      \\
$\beta$-20   & 4.00  & 0.94  &  &  $\alpha$-20 & 2.44 & 0.38      \\
$\beta$-22   & 4.00  & 0.88  &  &  $\alpha$-4  & 2.44 & 1.35    \\
$\alpha$-25  & 3.89  & 0.84  &  &  $\alpha$-2  & 2.44 & 0.84      \\
$\alpha$-12  & 3.89  & 0.84  &  &  $\alpha$-5  & 2.56 & 0.51      \\
$\alpha$-16  & 3.89  & 0.84  &  &  $\alpha$-11 & 2.67 & 1.15    \\
\bottomrule
\end{tabular}
\caption{Expert ranking of top and bottom 10 designs}
\label{tab:expert_ranking}
\end{table}

\begin{table*}[ht]
\centering
% \fontsize{7pt}{8pt}\selectfont
% \resizebox{1.8\columnwidth}{!}{
\begin{tabular}{lcccccllllccccc} 
\toprule
\multicolumn{6}{c}{\textbf{Participants Preference} ($\Delta < 0$)}  & \multicolumn{1}{c}{} & \multicolumn{1}{c}{} & \multicolumn{1}{c}{} & \multicolumn{6}{c}{\textbf{Expert Preference} ($\Delta > 0$)}  \\ 
\cline{1-6}\cmidrule(l){10-15}
\textbf{ID} & \textbf{Exp} & \textbf{Work} & \textbf{E.R.} & \textbf{W.R.} & \textbf{$\Delta$} &  &  &  & \textbf{ID} & \textbf{Exp} & \textbf{Work} & \textbf{E.R.} & \textbf{W.R.} & \textbf{$\Delta$}  \\ 
\cmidrule{1-6}\cmidrule{10-15}
$\alpha$-7  & 2.22  & 7.06  & 57  & 4  & $-53$  &   &    &  & $\beta$-12  & 4.33  & 6.13   & 4  & 50   & $+46$  \\
$\alpha$-10 & 2.67   & 7.12 & 50 & 2   & $-48$  &  &  &  & $\alpha$-15 & 4.67 & 6.13  & 3   & 49  & $+46$  \\
$\alpha$-5  & 2.56 & 6.93  & 52  & 6   & $-46$  & & & & $\alpha$-9  & 3.78  & 5.70 & 13  & 56 & $+43$ \\
$\alpha$-1  & 2.89 & 7.29 & 46 & 1  & $-45$  &   &  &  & $\beta$-26  & 4.11  & 6.21  & 5 & 44 & $+39$ \\
$\alpha$-11 & 2.67 & 6.96 & 50 & 5 & $-45$ &  &  &  & $\alpha$-25 & 3.89 & 6.18  & 8  & 46 & $+38$ \\
$\alpha$-29 & 2.33 & 6.71 & 56   & 14 & $-42$   & &    &    & $\alpha$-3  & 3.56  & 5.65  & 22   & 57 & $+35$ \\
$\alpha$-19 & 2.22         & 6.66          & 57            & 19            & $-38$             &                      &                      &                      & $\beta$-23  & 4.78         & 6.39          & 1             & 35            & $+34$              \\
$\beta$-11  & 3.22         & 6.76          & 39            & 9             & $-30$             &                      &                      &                      & $\beta$-22  & 4.00         & 6.33          & 6             & 37            & $+31$              \\
$\alpha$-21 & 2.67         & 6.65          & 49            & 20            & $-29$             &                      &                      &                      & $\beta$-4   & 4.67         & 6.46          & 2             & 32            & $+30$              \\
$\alpha$-32 & 3.11         & 6.70          & 42            & 15            & $-27$             &                      &                      &                      & $\beta$-3   & 3.44         & 5.94          & 25            & 52            & $+27$   \\
\bottomrule
\multicolumn{15}{l}{\textit{Note: Exp = Expert Mean, Work = Workshop Mean, E.R. = Expert Rank, W.R. = Workshop Rank.}} 
\end{tabular}
% }
\caption{Comparison of Rank Discrepancies: Participants Ranked Higher (Left) vs. Experts Ranked Higher (Right)}
\label{tab:rank_discrepancies}
\end{table*}

Furthermore, we analysed 237 expert comments that reveal complementary selection pressures: comfort/fluency (co-design participants) versus distinctiveness/robustness (experts). We found that a practical synthesis: retain participant-preferred, short, saccade-friendly geometries, but (i) add a brief arming dwell and a minimal commit (one blink or tap-stroke); (ii) enforce blink count windows and minimum inter-blink intervals; (iii) add angular tolerances and avoid reading-aligned single sweeps; (iv) calibrate aperture/thresholds where used; and (v) provide lightweight feedback during dwell. This keeps the interaction fast and learnable while meeting the experts’ requirements for recognizability and safety. Participants from co-design workshops consistently elevated single, short strokes (e.g., \texttt{$\alpha-1$}, \texttt{$\alpha-10$}, \texttt{$\alpha-7$}, \texttt{$\alpha-11$}, \texttt{$\alpha-5$}) because these feel effortless, quick, and easy to remember at first encounter. In contrast, expert noted that such cardinal, one-pass sweeps are hard to distinguish from natural viewing behaviour (reading, scanning) and thus invite unintended activation. Experts explicitly asked for a distinctive signature---for example short, composed patterns (L–R–L or U–D–U), or a brief arming dwell followed by a minimal commit action (blink or tap-stroke)---to separate looking from acting and mitigate the Midas Touch. They also recommended lightweight feedback (e.g., a dwell progress indicator at the edge) to make the arm/commit boundary legible. We highlighted that \textbf{participants prioritised fluency and immediate comfort; experts foregrounded distinctiveness and false-positive control}.

In particular, co-design participants from phase 1 often preferred designs that encode an explicit arm$\rightarrow$act grammar and use ``native ocular primitives''. For instance, blink-count patterns and two-stage gestures like, $\alpha-15$, $\alpha-17$, $\beta-11$, etc. They emphasised that multi-blink SOS and triple-blink exit gestures were judged as clear and deliberate, provided that blink windows and inter-blink intervals were properly enforced. In contrast, experts gave such patterns only mid-tier rankings, describing them as heavier or finicky compared to a single stroke---an attitude also reflected in critiques of longer sequences (\texttt{$\beta-23$}: ``too long and hard to remember the exact number''). However, we found that all experts praised the metaphor of \texttt{$\beta-26$} (aperture-based brightness), yet flagged the need for calibration and thresholds (codebook: Calibration \& Personalisation); while co-design participants appeared sceptical about comfort and feasibility without a working system. Conversely, letterforms with weak modality fit (e.g., \texttt{$\beta-21$}, ``Z'' for zoom) were criticised by experts for fatigue and weak mapping even when co-design participants found them memorable. We highlighted that \textbf{experts rewarded compositional grammar and modality-fit; participants were cautious about effort and unfamiliarity}.

\begin{tapscallout}
\noindent\textbf{Findings for RQ2:} While co-design participants favoured gestures that felt fluent, intuitive, and easy to perform---often preferring short, single-stroke designs---experts prioritized gestures that minimized false positives, aligned with natural ocular mechanics, and followed a clear compositional structure. Gestures lacking a distinctive signal or relying on overextended metaphors (e.g., pinch-to-zoom) were penalized, whereas designs with semantic clarity, reliable detection cues (e.g., blink-to-commit), and personalized calibration (e.g., squint-to-dim) were endorsed. Notably, experts and participants often diverged in their preferences, revealing a tension between immediate usability and long-term reliability. However, the findings also suggest that combining participant-preferred saccade-friendly shapes with expert-endorsed safeguards (e.g., timing constraints, compositional grammar, and lightweight feedback) to create gestures that are both intuitive and robust.
\end{tapscallout}

% \begin{tcolorbox}[colback=green!10, colframe=black!20, boxrule=0.5pt, sharp corners=southwest]
% \textbf{Findings for RQ2:} While co-design participants favoured gestures that felt fluent, intuitive, and easy to perform---often preferring short, single-stroke designs---experts prioritized gestures that minimized false positives, aligned with natural ocular mechanics, and followed a clear compositional structure. Gestures lacking a distinctive signal or relying on overextended metaphors (e.g., pinch-to-zoom) were penalized, whereas designs with semantic clarity, reliable detection cues (e.g., blink-to-commit), and personalized calibration (e.g., squint-to-dim) were endorsed. Notably, experts and participants often diverged in their preferences, revealing a tension between immediate usability and long-term reliability. However, the findings also suggest that combining participant-preferred saccade-friendly shapes with expert-endorsed safeguards (e.g., timing constraints, compositional grammar, and lightweight feedback) to create gestures that are both intuitive and robust.
% \end{tcolorbox}

\section{Discussion}
Our two-phase study, combining user-led co-design with formal expert evaluation, provides the unique lens through which to understand the creation of a practical and intuitive gaze gesture designs. The findings from Phase~1 reveal users' strong reliance on metaphor and existing interaction paradigms, while Phase~2 provides a critical filter for ergonomic and technical feasibility. By synthesizing these two perspectives, we move beyond a simple list of gestures, discussion the understanding gap between intuitive design and expert evaluation, and more for designing effective gaze interactions. We also reflect on the value of our two-phase methodology for gesture elicitation.

\subsection{The Gulf Between Intuition and Implementation: Deconstructing the Expert-Novice Gap}

Our findings reveal a fundamental and instructive disconnect between the gaze gestures non-experts intuitively design and those that experts deem viable. This gap is not a failure of user creativity but rather a reflection of deeply ingrained mental models shaped by other interaction modalities, primarily touch. Deconstructing this gap reveals three core tensions rooted in the distinct signal, biomechanical, and cognitive properties of the human eye. Understanding these tensions is critical for designing the next generation of intuitive and robust gaze interfaces.

\subsubsection{The Flawed Metaphor: The Eye as a Finger}
Lay participants consistently approached the design task by mapping their extensive experience with touchscreens directly onto gaze. They conceptualized the eye as a finger that could swipe, tap, and draw. While this provides a valuable starting point for metaphorical design, it collapses under scrutiny because it ignores the fundamental difference in the nature of the input signal.  

\textit{Touch is discrete; Gaze is continuous.} A touch interaction has a clear beginning (\texttt{touch-down}) and end (\texttt{touch-up}). This provides the system with unambiguous event markers to start and stop interpreting a gesture. In contrast, eye movement is an uninterrupted stream of data~\cite{land2009vision}. There is no inherent off state. This continuous nature is the source of the classic Midas Touch problem~\cite{jacob1990you, velichkovsky1997towards}: if looking is acting, how does a user simply look at something without activating it?  This is the reason why prior researchers prioritised gaze-gesture taxonomies and recognition performance studies that typically start from expert-defined primitives and then optimise accuracy of classifiers or recognizers~\cite{wobbrock2008longitudinal, porta2008eye, li2017gazture, Chen2021GazeGestureGCN, bature2023boosted, li2021evaluating, rozado2012gliding, chen2019gaze}. Non-experts intuitively stumbled upon this problem, leading them to independently invent solutions like \texttt{dwell time} and \texttt{blinking}. These mechanisms serve as a pseudo-commit signal, creating the discrete start/stop points that gaze naturally lacks. Experts validated and endorsed this approach, recognizing it as the core principle of \textit{compositionality} (e.g., dwell-to-arm, blink-to-fire) that separates observation from intention, a technique also explored in prior work~\cite{ramirez2021gaze+}. Conceptually, this two-stage grammar mirrors key aspects of direct manipulation, this is explicit control, continuous feedback, and reversible actions (i.e., the ability to cancel an 'armed' action)~\cite{shneiderman1983direct}, but reinterprets them for a modality where the signal is always ``on''.

\begin{tapscallout}
\noindent\textbf{Lesson to Learn:} 
Gaze is not a direct replacement for touch; it's a distinct modality requiring explicit mechanisms to manage its continuous signal. Effective design must provide users with a clear grammar to differentiate passive looking from active commanding.
\end{tapscallout}

% \begin{tcolorbox}[colback=green!10, colframe=black!20, boxrule=0.5pt, sharp corners=southwest]
% \textbf{Lesson to Learn:} 
% Gaze is not a direct replacement for touch; it's a distinct modality requiring explicit mechanisms to manage its continuous signal. Effective design must provide users with a clear grammar to differentiate passive looking from active commanding.
% \end{tcolorbox}

\subsubsection{The Physical Reality: Biomechanics over Aesthetics}
The second major point of divergence was the clash between visually appealing gesture shapes and the biomechanical constraints of the ocular-motor system. %\textit{Saccades and Smooth Pursuit.} 
Participants often designed visually elegant gestures like circles, spirals, and smooth curves, assuming the eye could trace them as easily as a finger. However, experts consistently rejected these designs because human eye movement is primarily \texttt{saccadic}. The eye's musculature, particularly the four rectus muscles, is optimized for rapid, ballistic jumps between points in straight lines~\cite{kandell2021principles, duong2008neuro}. Tracing a curve without a moving target to follow (smooth pursuit) requires immense concentration and the coordinated firing of multiple muscle groups (including the oblique muscles)~\cite{herrera2019cranial}. This leads to rapid fatigue and imprecision, as well as \textit{cognitive and muscular load}, explaining why experts favoured gestures composed of simple cardinal (horizontal/vertical) strokes.

\begin{tapscallout}
\noindent\textbf{Lesson to Learn:} 
Effective gaze gestures must prioritize ergonomic simplicity by aligning with the eye's natural in movements, i.e. saccadic (ballistic) motion. Physiological comfort and low cognitive load must trump aesthetic complexity, especially for frequently used commands.
\end{tapscallout}

% \begin{tcolorbox}[colback=green!10, colframe=black!20, boxrule=0.5pt, sharp corners=southwest]
% \textbf{Lesson to Learn:} 
% Effective gaze gestures must prioritize ergonomic simplicity by aligning with the eye's natural in movements, i.e. saccadic (ballistic) motion. Physiological comfort and low cognitive load must trump aesthetic complexity, especially for frequently used commands.
% \end{tcolorbox}

\subsubsection{The Cognitive Tension: The Intentional Eye vs. The Reflexive Eye}
Finally, the most subtle gap lies in the understanding of cognitive control. The novice assumption is that where we look is a direct and pure reflection of our conscious will. However, neuroscience shows that eye movements are governed by two competing systems~\cite{land2009vision, zelinsky2005role}.

\textit{Top-Down vs. Bottom-Up Control.} {Top-down control} is conscious and goal-directed~\cite{glaholt2010evidence}. {Bottom-up control}, in contrast, is reflexive and stimulus-driven~\cite{peters2005components}. This latter system is an evolutionary survival mechanism that operates largely outside our conscious control~\cite{zelinsky2005role}.

\textit{Designing for Intent.} This dual-control system is a major source of error in gaze interaction. A bottom-up, reflexive glance can easily be misinterpreted by a system as a top-down, intentional command. This is precisely why the \texttt{compositional gaze grammar} (dwell + action) is so critical. The initial dwell phase acts as a filter, confirming that the user is engaging in sustained, top-down attention before the system becomes receptive to an action command. It creates a moment for conscious intent to be established and validated.

\begin{tapscallout}
\noindent\textbf{Lesson to Learn:} 
Interaction design must account for the dual nature of gaze in information capture and output. A robust system never assumes that looking equals intention. It must provide an unambiguous structure that allows users to explicitly declare their commands, effectively filtering out reflexive, bottom-up glances.
\end{tapscallout}

% \begin{tcolorbox}[colback=green!10, colframe=black!20, boxrule=0.5pt, sharp corners=southwest]
% \textbf{Lesson to Learn:} 
% Interaction design must account for the dual nature of gaze in information capture and output. A robust system never assumes that looking equals intention. It must provide an unambiguous structure that allows users to explicitly declare their commands, effectively filtering out reflexive, bottom-up glances.
% \end{tcolorbox}

\subsubsection{From Gap to Synergy}
Crucially, our study demonstrates that this expert-novice gap is not insurmountable. Through a scaffolded co-design process, non-expert participants began to converge on principles long-established by researchers, such as the use of dwells~\cite{lei2023dynamicread, Namnakani2023CompareGaze}, gaze~+~blink/wink combinations~\cite{ramirez2021gaze+}, and simple gaze gesture~\cite{lei2023dynamicread, Namnakani2023CompareGaze, wobbrock2008longitudinal, porta2008eye}. This suggests that the gap is one of knowledge, not innate capability. \textit{Step 1 (Instruction Kick Off)} listen and learn the knowledge of the nature of eye movements and some example of gaze interaction projects. \textit{Step 2 (User Co-Design)} taps into the rich, semantic, and metaphorical understanding of users to ensure gestures are {intuitive and learnable}. \textit{Step 3 (Group Review and Discussion)} applies a critical filter based on physiological, cognitive, and technical constraints to ensure the gestures are {robust, comfortable, and reliable}.

\subsection{From Findings to Framework: Four Principles for User-Centric Gaze Gesture Design}

Synthesizing the insights from both Phase 1 and Phase 2 allows us to move beyond a mere list of gestures to a set of foundational principles. These principles represent the solutions to key design tensions that arose when user intuition collides with the realities of ergonomics and technology. They provide a robust framework for creating gaze interaction vocabularies that are not only intuitive but also reliable, comfortable, and safe. Ultimately, this framework serves to guide the design of gaze manipulation commands and scalable gaze interactions, laying the necessary groundwork for rigorous practical testing and real-world implementation.

\subsubsection{Principle 1: Establish a Clear Grammar of Intent}
\begin{figure}[ht]
    \centering
    \includegraphics[width=0.95\linewidth]{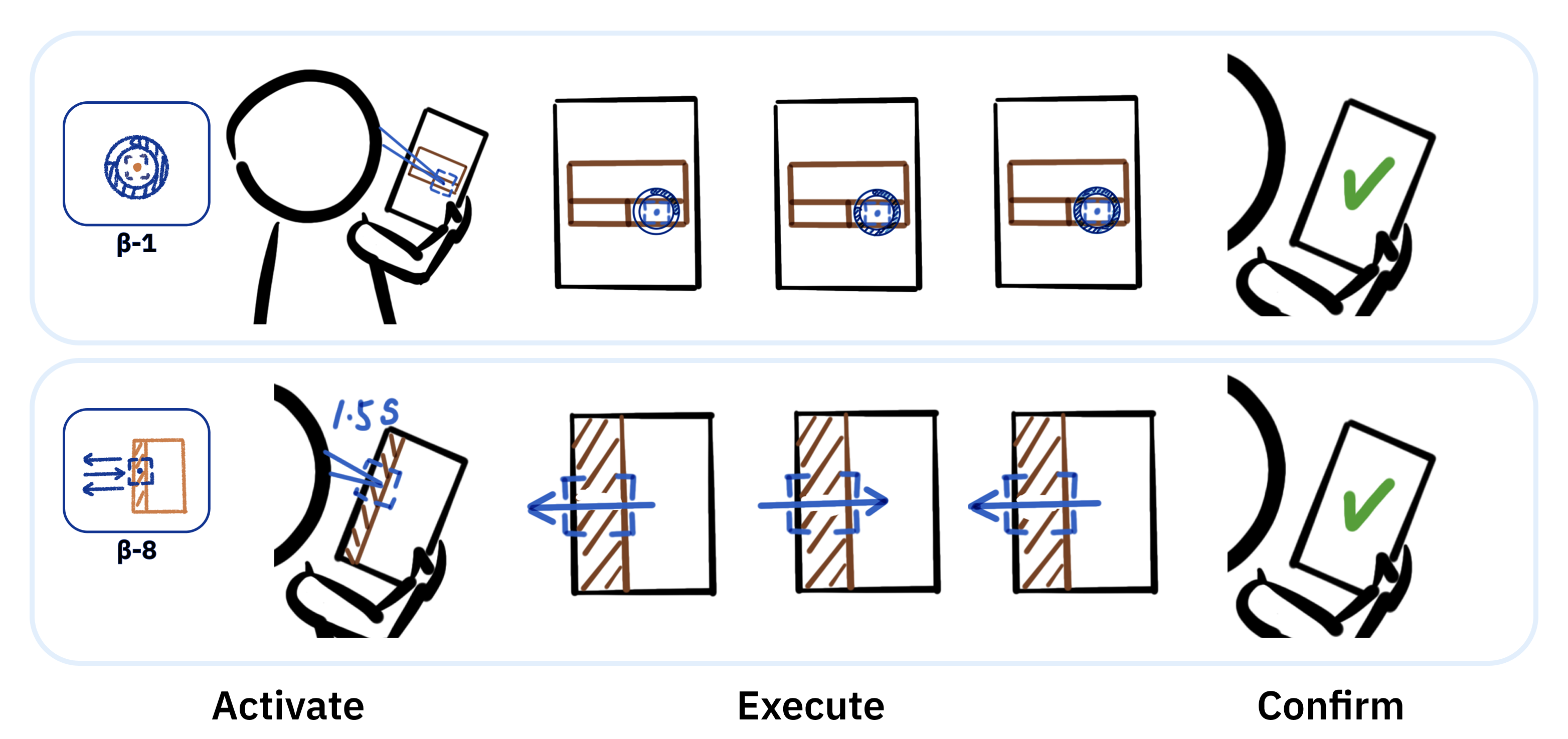}
    \caption{Schematic of the compositional gaze grammar used for command decoding: \emph{Activate} (a short dwell arms the interface), \emph{Execute} (the user performs a gaze gesture or blink to execute the command, reducing accidental triggers).}
    \Description{A three-stage interaction cycle for gaze input. The diagram shows an Activate step that arms input via a brief dwell, followed by an Action step where the user executes a gaze gesture (or blink) to confirm the command.}
    \label{fig:decode_grammer}
\end{figure}

The most significant challenge in gaze interaction is distinguishing passive observation from active command, the Midas Touch problem~\cite{jacob1990you}. Our study showed that both non-experts and experts converged on the same powerful solution: a two-stage interaction syntax. Users organically designed gestures with an \textit{activate-then-confirm} structure to feel in control (S1-GC-P11), a pattern experts unanimously validated as a fundamental prerequisite for reliability (S2-EP1, EP2, EP3). This moves beyond treating gestures as monolithic paths and establishes them as compound statements, see Figure~\ref{fig:decode_grammer}.

\begin{tapsbox}[blue!10]
\noindent\textbf{Design Takeaway 1: Structure gaze commands as an \emph{activate-then-confirm} two-part sequence.} An activation primitive (like a dwell) signals intent and prepares the system, while a subsequent confirmation primitive (a scanpath or a discrete blink) executes the command. This compositional gaze grammar is the most robust method for preventing accidental triggers.
\end{tapsbox}

% \begin{tcolorbox}[colback=blue!10, colframe=black!20, boxrule=0.5pt, sharp corners=southwest]
% \textbf{Design Takeaway 1: Structure gaze commands as an \emph{activate-then-confirm} two-part sequence.} An activation primitive (like a dwell) signals intent and prepares the system, while a subsequent confirmation primitive (a scanpath or a discrete blink) executes the command. This compositional gaze grammar is the most robust method for preventing accidental triggers.
% \end{tcolorbox}

\subsubsection{Principle 2: Prioritize Physiological Reality over Naive Metaphor}
Users naturally draw on familiar metaphors from touch interfaces, proposing gestures like circles for copy or pinching motions for zoom ($\beta-18$). While this grounding is excellent for learnability, our expert evaluation provided a critical filter: metaphors must respect the body's physical constraints. Experts consistently rejected complex curves and spirals, noting that the eye's musculature is optimized for rapid, linear \textbf{saccades}, not the smooth pursuit required to trace shapes without a moving guide (clockwise or counter-clockwise gestures, i.e. $\alpha-12$, $\alpha-33$, $\beta-14$ and $\beta-15$ in Fig.~\ref{fig:merge_example}). A gesture that is intuitive in concept can be deeply uncomfortable and imprecise in practice if it fights against the eye's natural mechanics.

\begin{tapsbox}[blue!10]
\noindent\textbf{Design Takeaway 2: Favour gestures composed of linear, saccade-friendly segments.} Prioritize straight lines and simple angles over curves or complex paths. The long-term usability of a gesture depends more on its ergonomic comfort and low physical effort than on the novelty or directness of its metaphor.
\end{tapsbox}

% \begin{tcolorbox}[colback=blue!10, colframe=black!20, boxrule=0.5pt, sharp corners=southwest]
% \textbf{Design Takeaway 2: Favour gestures composed of linear, saccade-friendly segments.} Prioritize straight lines and simple angles over curves or complex paths. The long-term usability of a gesture depends more on its ergonomic comfort and low physical effort than on the novelty or directness of its metaphor.
% \end{tcolorbox}

\subsubsection{Principle 3: Match Interaction Cost to Operational Consequence}
\begin{figure}[ht]
  \centering
  \includegraphics[width=\linewidth]{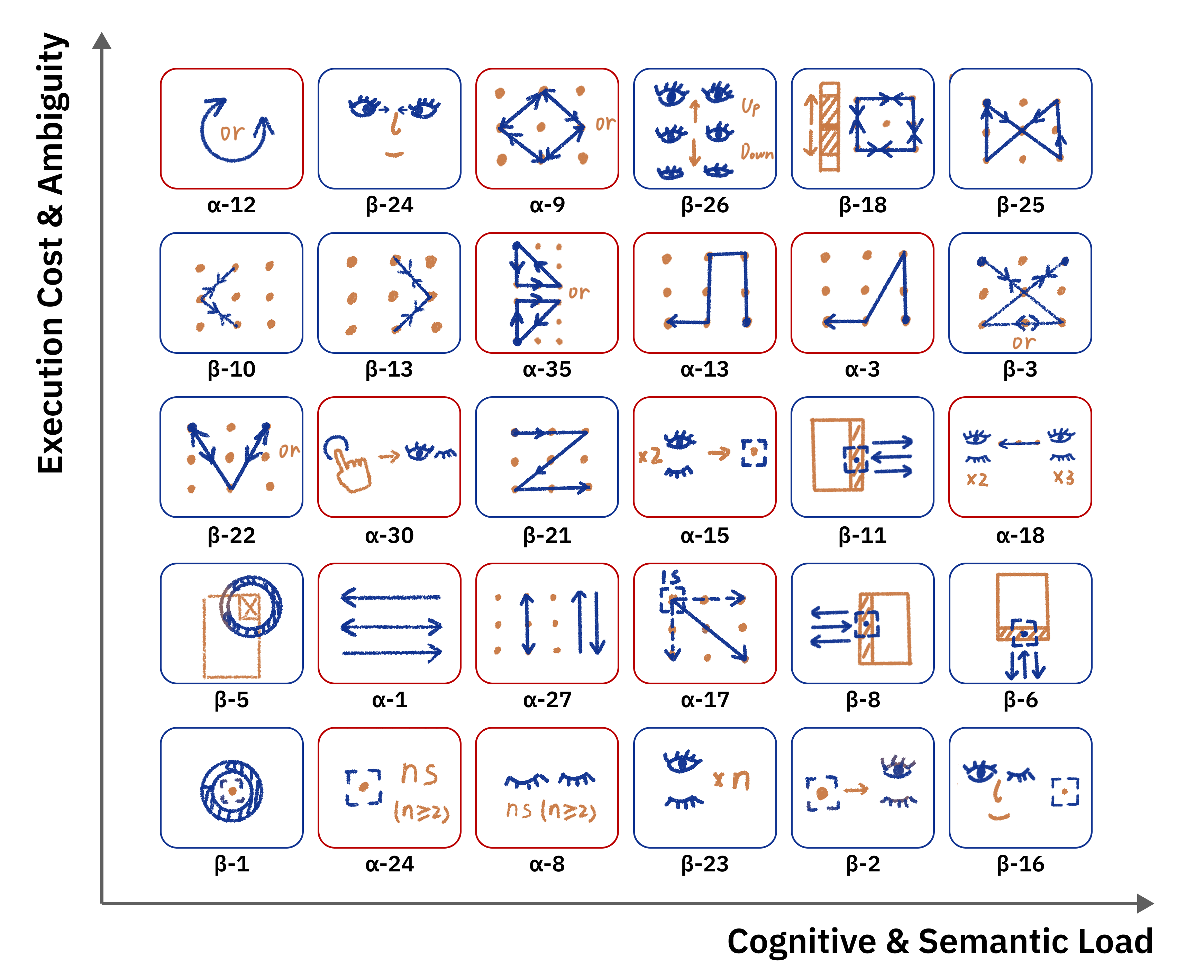}
  \caption{The design trade-off space for the final gesture set. Gestures are mapped according to their cognitive and semantic load ($x$-axis) versus their physical execution cost and ambiguity ($y$-axis). The distribution illustrates the principle of proportional complexity: simple(aligned with eye movements), low-cost (low motion complexity and easy to memorise) gestures (bottom-left) are ideal for frequent commands, while more complex, high-cost gestures (top-right) serve as intentional safeguards for critical actions.
  Gestures $\alpha-26$ and $\alpha-34$ are excluded from this mapping as their complexity is either user-dependent or a variant of $\beta-5$ gesture. 
  }
  \Description{A two-dimensional scatter plot mapping the final gesture set. The Y-axis represents "Execution Cost \& Ambiguity" (low to high), and the X-axis represents "Cognitive \& Semantic Load" (low to high). Gestures are distributed across the quadrants. Simple gestures like directional swipes are in the bottom-left (low cost, low load). Complex gestures like multi-step shapes or specific blink sequences are in the top-right (high cost, high load).}
  \label{fig:phase2_tradeoff}
\end{figure}

A core tension emerged between the desire for speed and the need for safety. While participants often favored the simplest and fastest gestures, experts introduced the crucial concept of \textbf{proportional complexity}. They argued that high-stakes, irreversible commands (e.g., Shut Down, Delete) should be \textit{intentionally} more complex to execute. To make this trade-off more tangible, we mapped representative gestures onto a two-dimensional design space contrasting \emph{Execution Cost \& Ambiguity} with \emph{Cognitive \& Semantic Load} (Fig.~\ref{fig:phase2_tradeoff}). As shown, gestures such as $\beta-25$ occupy the upper-right quadrant, where added cognitive and physical effort functions as a natural safeguard against accidental activation, transforming a potential usability flaw (i.e., being cumbersome) into a valuable safety feature. This was also reflected in users' designs for emergency SOS gestures, which involved deliberate, unnatural patterns like rapid multi-blink sequences and inward rotation of both eyes ($\beta-23$ and $\beta-24$ in Fig.~\ref{fig:phase1_by_command}).

\begin{tapsbox}[blue!10]
\noindent\textbf{Design Takeaway 3: Calibrate gesture complexity against command risk.} Use simple, low-effort gestures for frequent, low-consequence actions (e.g., next/previous). Reserve more complex, multi-step, or physically deliberate gestures for critical, high-consequence actions to minimize the chance of catastrophic errors.
\end{tapsbox}

% \begin{tcolorbox}[colback=blue!10, colframe=black!20, boxrule=0.5pt, sharp corners=southwest]
% \textbf{Design Takeaway 3: Calibrate gesture complexity against command risk.} Use simple, low-effort gestures for frequent, low-consequence actions (e.g., next/previous). Reserve more complex, multi-step, or physically deliberate gestures for critical, high-consequence actions to minimize the chance of catastrophic errors.
% \end{tcolorbox}

\subsubsection{Principle 4: Anchor the gaze gestures in Familiar Contexts}

Our findings show that the most successful and learnable gestures were anchored in users' existing knowledge. This anchoring happened in two key ways. First, \textbf{metaphorical anchoring} involved transferring established conventions from GUI and touch interfaces (e.g., drawing an `X' to close, swiping to go back) ($\alpha-34$, $\beta-3$, $\beta-10$ and $\beta-25$ in Fig.~\ref{fig:final_gestures}). Second, \textbf{spatial anchoring} involved tying gestures to stable landmarks on the device, such as the screen edges and corners ($\alpha-34$, $\beta-5$, $\beta-6$ etc. in Fig.~\ref{fig:final_gestures}). These edge-anchored gestures were praised by users for being efficient and non-intrusive ($\alpha-34$, $\beta-6$, $\beta-8$ and $\beta-11$ in Fig.~\ref{fig:final_gestures}). This principle suggests that a foundational gaze syntax should not try to reinvent all interaction paradigms, but rather provide a gentle bridge from the familiar to the new.

\begin{tapsbox}[blue!10]
\noindent\textbf{Design Takeaway 4: Ground gestures in users' existing mental models.} Anchor them metaphorically to familiar interactions and spatially to stable landmarks like screen edges. This dramatically lowers the initial learning curve and makes the system feel intuitive from the start.
\end{tapsbox}

% \begin{tcolorbox}[colback=blue!10, colframe=black!20, boxrule=0.5pt, sharp corners=southwest]
% \textbf{Design Takeaway 4: Ground gestures in users' existing mental models.} Anchor them metaphorically to familiar interactions and spatially to stable landmarks like screen edges. This dramatically lowers the initial learning curve and makes the system feel intuitive from the start.
% \end{tcolorbox}

As a sanity check on the generality of this gesture set, Appendix~\ref{app:prior_work_comparison} presents a post-hoc mapping between our user-generated gestures and representative gaze gesture forms reported in prior work. This comparison shows that participants, without seeing any of these earlier designs, independently rediscovered many of the same gesture families (e.g., directional flicks, crosses, simple zig-zags), supporting the claim that our four principles capture stable regularities in how people appropriate gaze for command input rather than idiosyncratic artefacts of our study. At the same time, the mapping also surfaces mobile- and context-specific variants (e.g., edge-anchored flicks and emergency blink patterns) that extend existing desktop- and VR-focused vocabularies, pointing to opportunities for platform-adaptive gaze gesture sets.

\subsection{Reflections on the Two-Phase Co-Design Methodology}

Our two-phase methodology, combining bottom-up user creativity with top-down expert critique, proved essential for navigating the challenges of gaze interaction design. This process did more than generate and filter gestures; it revealed the synergy required to create interactions that are both intuitive and technically viable.

In Phase 1, non-expert participants demonstrated remarkable creativity, grounding their designs in semantic and metaphorical intuition. They drew on familiar interaction paradigms (e.g., using an ``X'' to mean close), proposed expressive somatic gestures (e.g., squinting to dim), and even designed context-aware commands like blink-based SOS signals. This shows that users, when given the opportunity, are capable of ideating rich, meaningful, and innovative interactions that go far beyond simple command-and-control. Crucially, our findings suggest that the knowledge gap for designing effective gaze interactions is not as wide as often assumed. With minimal scaffolding, a brief introduction to core concepts and a simple 9-point grid---participants were able to quickly grasp the modality's constraints and possibilities. They independently converged on foundational interaction patterns, such as the compositional \textit{activate-then-confirm} gaze grammar, that have long been established by domain experts. This indicates that the learning barrier for contributing to gaze design is surmountable, opening the door for more diverse voices in the design process.

The expert peer-review in Phase 2 did not stifle this creativity but rather channelled it. The experts provided the necessary physiological and technical guardrails, refining the user-generated concepts into a robust and comfortable set by eliminating gestures that would cause undue physical strain (e.g., complex curves) or that would be nearly impossible for a system to reliably distinguish from natural eye movements (e.g., simple, single swipes).

This two-phase process---where users define \textit{what} feels natural and experts refine \textit{how} it can work reliably---therefore provides more than just a framework for vocabulary development. It demonstrates that the barrier to entry for designing for gaze is lower than perceived. By empowering a wider range of designers, developers, and product managers with this collaborative model, we can accelerate the creation of innovative and inclusive gaze-enabled applications. This ensures that the benefits of this powerful, hands-free modality can reach a broader audience, particularly users with motor impairments who stand to gain the most.

\section{Limitations} \label{method_limitation}

As with many co-design studies, self-reported data may be affected by social desirability bias~\cite{nederhof1985methods}. Although triangulated with observational data, think-aloud notes, and facilitator insights, participants may have over- or under-emphasized certain design rationales or usability concerns. Despite scaffolding non-expert creativity through a 9-point grid and briefing materials, limited familiarity with gaze interaction may have constrained the realism or technical depth of some proposals. Participants were primarily recruited from university communities and mailing lists, resulting in a relatively homogeneous sample in terms of education and age. While this facilitated engagement and language proficiency, it may limit generalizability. Future studies should recruit more diverse users, including older adults, individuals with lower digital literacy, and those with accessibility needs. In Phase 2, expert review added critical evaluation of gesture feasibility and robustness, but the panel included only four experts, all with academic backgrounds in gaze interaction. While this ensured rigour, it may have narrowed the lens toward academic perspectives. Including practitioners, designers, or developers in future panels could offer more applied insights. Although AC-led discussions were structured to reduce dominance bias, group dynamics may still have shaped consensus. Lastly, the study focused on mobile gesture design using a predefined 9-point grid. While this aided comparison and implementation filtering, it may have constrained exploration of more fluid or unconventional forms (e.g., continuous paths, pressure-based gestures). 

Our evaluation focuses on expert judgement and workshop ratings rather than end-to-end performance with running recognisers. We do not, for example, report command-level error rates, throughput, or long-term memorability in deployed applications. As such, the gesture vocabularies should be treated as an expert-curated candidate set and a set of design heuristics rather than a fully optimised standard. Future work should explore more flexible frameworks, alternative platforms (e.g., desktop), and real-time prototyping to assess viability in situ, i.e. learnability, speed, error resilience, and social acceptability across diverse user groups, devices, and usage contexts.

\section{Conclusion}
In this study, we addressed the urgent need for a user-centred gaze gesture command set as eye-tracking becomes ubiquitous in mobile devices. Through a two-phase methodology that combined participatory co-design workshops with 20 non-experts and a structured peer-review by four domain experts, we developed and validated a set of 32 gestures. Our findings reveal a crucial synergy: non-expert users contribute intuitive, metaphor-driven designs and spontaneously invent core interaction structures---most notably a compositional grammar of activate-then-confirm (e.g., dwell-then-blink)---to ensure intentionality and mitigate the Midas Touch problem. Crucially, this co-design success demonstrates that non-specialists can master the gaze interaction design process with minimal scaffolding, lowering the barrier for its adoption and innovation across broader domains. The subsequent expert evaluation provides an essential filter for physiological and technical feasibility, prioritizing ergonomically sound, saccade-friendly gestures with low ambiguity to ensure reliable machine recognition. By synthesizing these perspectives, our work makes three key contributions. First, we provide an empirically derived and expert-validated set of gaze gestures optimized for mobile use. Second, we formalize the principle of compositionality as a foundational syntax for designing reliable gaze interactions. Third, we distil our findings into actionable design principles that bridge the gap between user intuition and implementation reality. This research offers both a practical gesture set and a collaborative design framework to guide the development of more intuitive, accessible, and robust gaze-enabled interfaces. Future work should focus on implementing this gesture set in live systems to evaluate its performance with diverse users in real-world conditions.

% ======== Main Content ========%

\begin{acks}
We thank all participants for their time and contributions. We also thank the CHI reviewers for their constructive feedback, which helped improve the clarity and presentation of this work. Finally, we are grateful to our colleagues for helpful discussions and support throughout the project.
\end{acks}

\bibliographystyle{ACM-Reference-Format}
\bibliography{reference}

\clearpage
\appendix

{
% \color{blue}
\section{Comparison of User-Generated Gestures with Prior Literature}\label{app:prior_work_comparison}

Figure~\ref{fig:alignment} compares the gaze gestures produced in our co-design study with representative similar gesture forms reported in prior work. The examples shown are representative rather than exhaustive, and the correspondences reflect approximate morphological similarity rather than strict one-to-one trajectory alignment. Our goal is not to claim direct equivalence but to highlight how participants spontaneously reproduced shapes that resemble well-established gaze interaction patterns.

For clarity, the literature sources associated with each group of sub-figures are
listed below:

\begin{itemize}
  \item (a--f, n, ab, ah, ai, aq, as, at, au, ay, be):~\cite{wobbrock2008longitudinal}
  \item (g, h, o, r, w, y, ac, ak--an, aw, az, bb--bd):~\cite{porta2008eye}
  \item (s, u, v, ad):~\cite{istance2017supporting}
  \item (t, z, ae, aj):~\cite{drewes2007interacting}
  \item (l):~\cite{bature2023boosted}
  \item (av, ax, bf):~\cite{li2021evaluating}
  \item (ao, ap, ar):~\cite{rozado2012gliding}
  \item (i, af):~\cite{li2017gazture}
  \item (j, k, ag, ba):~\cite{chen2019gaze}
  \item (aa, bg, bk):~\cite{lei2023dynamicread}
  \item (bn, bo):~\cite{ramirez2021gaze+}
  \item (bu):~\cite{pfeuffer2016gaze}
  \item (bv):~\cite{rivu2020gaze}
  \item (m, bl, bm):~\cite{vidal2013pursuits}
  \item (bh):~\cite{Namnakani2023CompareGaze}
  \item (bi):~\cite{mollenbach2013eye}
  \item (bj):~\cite{fernandez2020gazewheels}
  \item (br):~\cite{ezzat2023blink}
  \item (bp):~\cite{brefczynski2011blink}
  \item (bq):~\cite{cheng2023twinkletwinkle}
  \item (bs, bt, bx, by):~\cite{rolff2025hands}
\end{itemize}

\begin{figure*}[t]
  \centering
  \includegraphics[width=\textwidth]{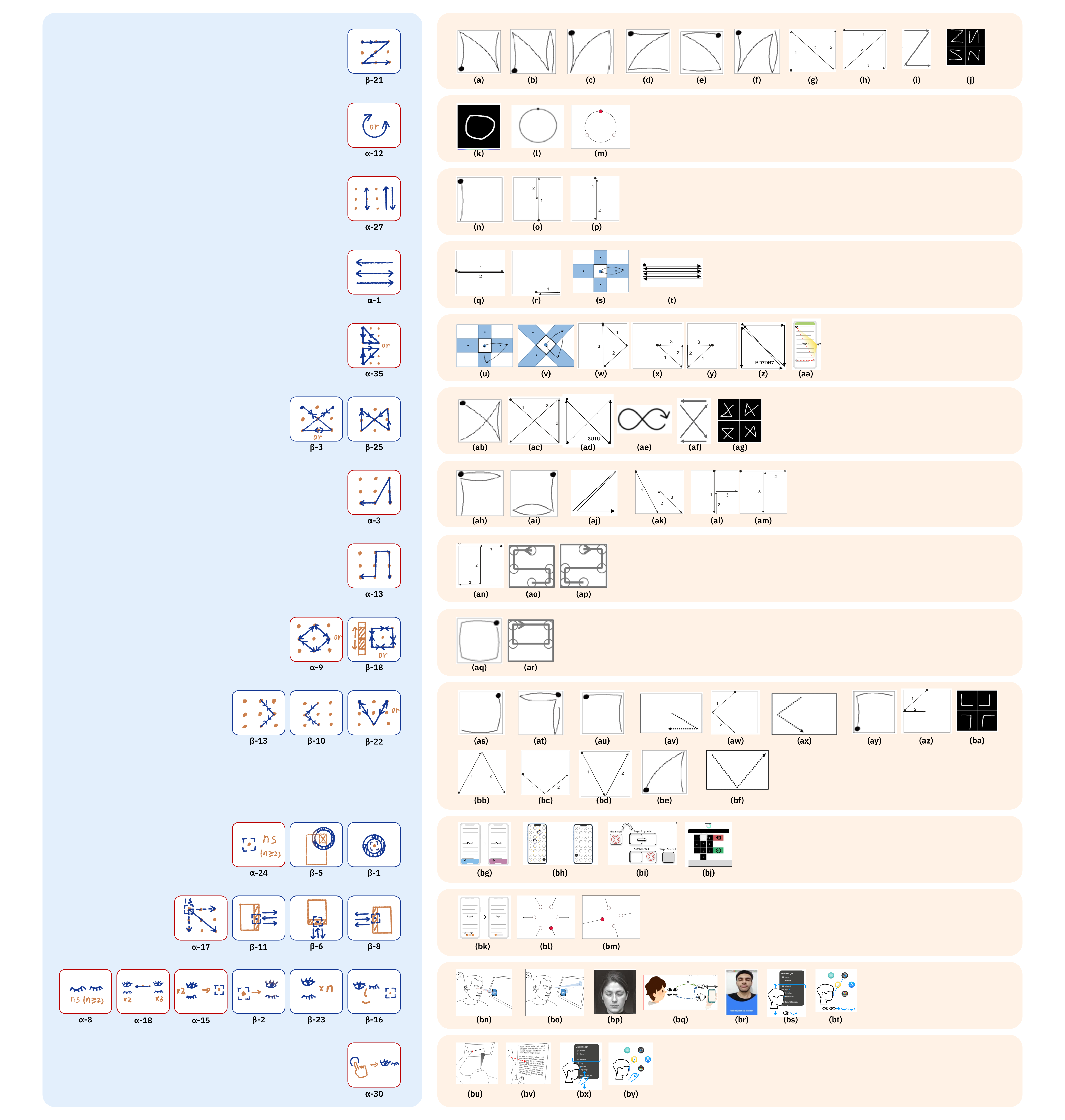}
  \caption{
  Overview of gesture-shape correspondences between user-generated gaze gestures and representative similar forms reported in prior literature. Each sub-figure (a--by) shows an approximate morphological similarity rather than exact trajectory alignment.}
  \Description{A comparison chart mapping user-generated gestures to similar forms found in prior academic literature. The chart is organized into a grid where each cell contains a user gesture next to a visually similar gesture from cited papers. The pairs show consistent patterns such as straight lines, corner triangles, and circular motions.}
  \label{fig:alignment}
\end{figure*}

% \begin{figure*}[t]
%   \centering
%   \includegraphics[width=0.65\textwidth]{figure/alignment_overview.png}
%   \caption{%
%   Mapping between gesture forms from representative prior work and the user-generated
%   gaze gestures in our co-design study. Each sub-figure (a--by) illustrates a gesture
%   shape previously reported in the literature and its approximate morphological
%   similarity to user-defined gestures. The examples are representative rather than
%   exhaustive, and the mapping reflects shape similarity rather than exact trajectory
%   alignment rather than exact trajectory correspondence.%
%   \textit{Illustrative references:}
%   (a--f, n, ab, ah--ai, aq, as--at, au, ay, be)~\cite{wobbrock2008longitudinal};
%   (g, h, o--r, w--y, ac, ak--an, aw, az, bb--bd)~\cite{porta2008eye};
%   (s, u--v, ad)~\cite{istance2017supporting};
%   (t, z, ae, aj)~\cite{drewes2007interacting};
%   (l)~\cite{bature2023boosted};
%   (av, ax, bf)~\cite{li2021evaluating};
%   (ao--ap, ar)~\cite{rozado2012gliding};
%   (i, af)~\cite{li2017gazture};
%   (j, k, ag, ba)~\cite{chen2019gaze};
%   (aa, bg, bk)~\cite{lei2023dynamicread};
%   (bn, bo)~\cite{ramirez2021gaze+};
%   (bu)~\cite{pfeuffer2016gaze};
%   (bv)~\cite{rivu2020gaze};
%   (m, bl--bm)~\cite{vidal2013pursuits};
%   (bh)~\cite{Namnakani2023CompareGaze};
%   (bi)~\cite{mollenbach2013eye};
%   (bj)~\cite{fernandez2020gazewheels};
%   (br)~\cite{ezzat2023blink};
%   (bp)~\cite{brefczynski2011blink};
%   (bq)~\cite{cheng2023twinkletwinkle};
%   (bs, bt, bx--by)~\cite{rolff2025hands}.}
%   \label{fig:alignment}
% \end{figure*}

\section{Gesture Icon Definitions}\label{app:gesture_icon_definition}

This appendix provides the final descriptive meanings of all gaze gesture icons used in our gesture library (Figure~\ref{fig:phase1_gestures_appendix}).

\begin{figure*}[!htbp]
  \centering
  \includegraphics[width=\linewidth]{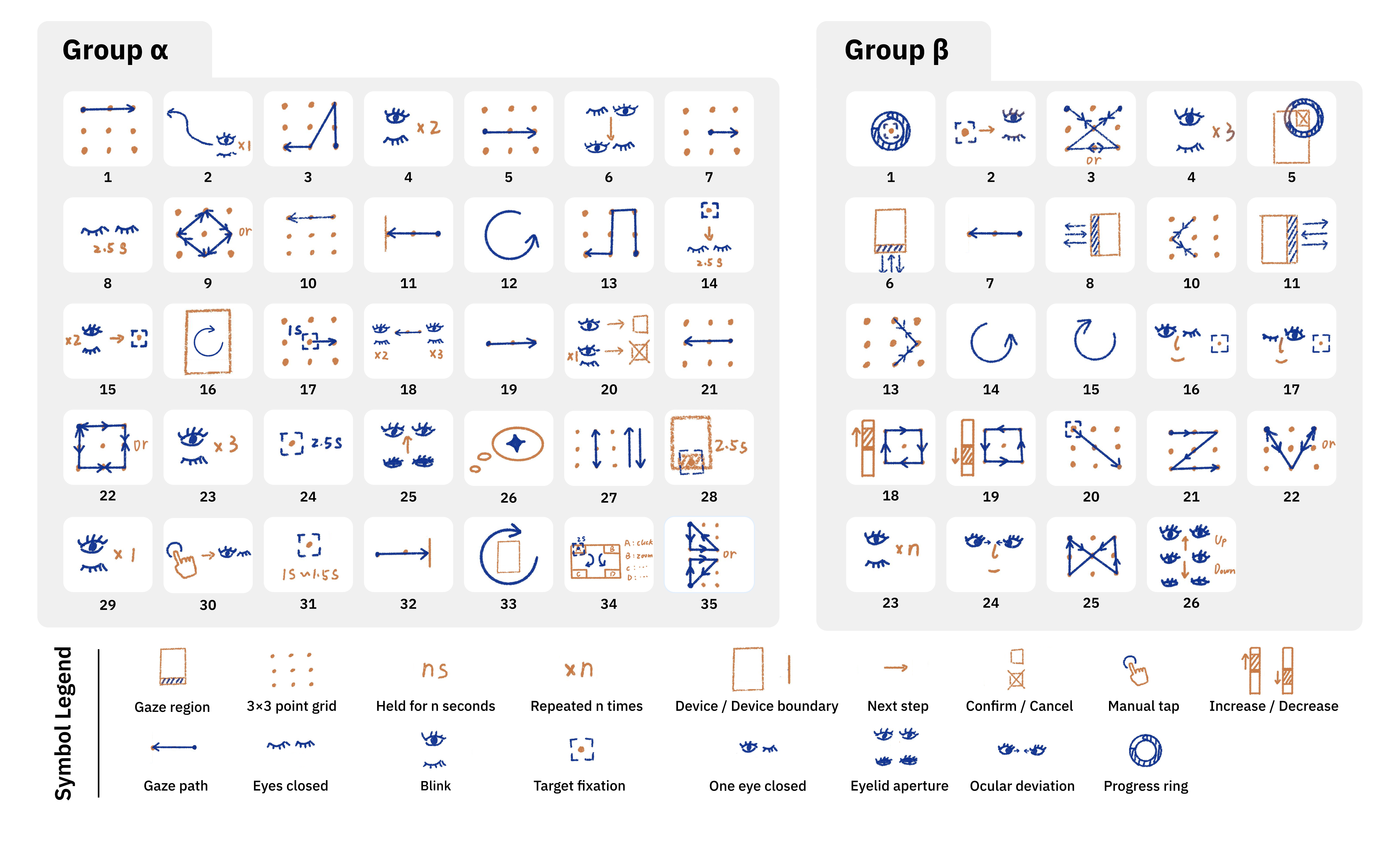}
  \caption{The complete set of 59 participant generated gaze gestures in standardized format.}
  \Description{In Appendix B, a large grid displaying 59 standardized icons of gaze gestures, divided into two sections: "Group Alpha" (gestures 1-35) and "Group Beta" (gestures 1-26). Each gesture is depicted using blue lines for gaze paths and orange symbols for interface elements or dwell times. A legend at the bottom explains the symbols, including: Gaze path (arrow), Eyes closed (lashes), Blink (open eye with lashes), Target fixation (brackets), One eye closed, Eyelid aperture, Ocular deviation, Progress ring, 3x3 point grid, Dwell time (ns), and Repetition (xn).}
  \label{fig:phase1_gestures_appendix}
\end{figure*}

\subsection*{Group $\alpha$}

\begin{enumerate}[label=\textbf{$\alpha$\arabic*}.]
    \item Horizontal movement from top-left to top-right.
    \item Blink once, then move diagonally from bottom-right to top-left along an arc.
    \item Follow a dot grid path: bottom-right $\rightarrow$ top-right $\rightarrow$ bottom-center $\rightarrow$ bottom-left.
    \item Blink twice consecutively.
    \item Horizontal movement from left to right.
    \item Open the right eye only, then the left eye only.
    \item Horizontal movement from the center to the right.
    \item Close both eyes for approximately 2.5 seconds.
    \item Cycle among four corner points (diamond trajectory, clockwise or counterclockwise).
    \item Horizontal movement from top-right to top-left.
    \item Horizontal movement from right to left toward the boundary.
    \item Circular gaze sweep in a counterclockwise direction.
    \item Move upward from bottom-right, then left, then down, then left (following a dot grid path).
    \item Fixate on a target, then close the eyes for approximately 2.5 seconds.
    \item Blink twice consecutively, then fixate on a target.
    \item Perform a clockwise circular sweep within the screen area.
    \item Fixate for 1 second, then move horizontally to the right.
    \item Blink three times $\rightarrow$ move right $\rightarrow$ blink twice.
    \item Horizontal straight-line movement from left to right.
    \item Fixation indicates ``select'', blinking indicates ``cancel''.
    \item Horizontal movement from right to left at the central height.
    \item Cycle among the four corners of a rectangle (clockwise or counterclockwise).
    \item Blink three times consecutively.
    \item Fixate on a target area for approximately 2.5 seconds.
    \item Gradually widen both eyes.
    \item User-defined action.
    \item Vertical movement upward and downward.
    \item Fixate for approximately 2.5 seconds on a fixed region near the edge of the phone.
    \item Blink once.
    \item Press a specific button and then blink once.
    \item Maintain fixation for 1--1.5 seconds to trigger an action.
    \item Horizontal movement from left to right toward the boundary.
    \item Perform a clockwise circular sweep outside the screen area.
    \item Fixate on one of the four screen corners for about 2 seconds, then naturally sweep downward.
    \item Follow a right-triangle contour (clockwise or counterclockwise).
\end{enumerate}

\subsection*{Group $\beta$}

\begin{enumerate}[label=\textbf{$\beta$\arabic*}.]
    \item[\textbf{$\beta$1}.] Continuous fixation on a target area, with a circular progress ring indicating activation.
    \item[\textbf{$\beta$2}.] Fixate on a point, then blink once.
    \item[\textbf{$\beta$3}.] Move in the sequence: top-left $\rightarrow$ bottom-right $\rightarrow$ bottom-left $\rightarrow$ top-right, or the reverse order.
    \item[\textbf{$\beta$4}.] Blink three times consecutively.
    \item[\textbf{$\beta$5}.] Fixation triggers an action, accompanied by a circular progress ring.
    \item[\textbf{$\beta$6}.] Sweep vertically downward from the top of the screen, then upward, and upward again.
    \item[\textbf{$\beta$7}.] Horizontal movement from right to left.
    \item[\textbf{$\beta$8}.] Sweep vertically from the left side $\rightarrow$ to the right $\rightarrow$ then diagonally upward to the left.
    \item[\textbf{$\beta$10}.] Follow a dot grid: top-center $\rightarrow$ left-center $\rightarrow$ bottom-center, then return along the same path repeatedly.
    \item[\textbf{$\beta$11}.] Sweep from the right side vertically outward $\rightarrow$ inward $\rightarrow$ farther inward.
    \item[\textbf{$\beta$13}.] Follow a dot grid: top-center $\rightarrow$ right-center $\rightarrow$ bottom-center, then return along the same path repeatedly.
    \item[\textbf{$\beta$14}.] Circular sweep in a counterclockwise direction.
    \item[\textbf{$\beta$15}.] Circular sweep in a clockwise direction.
    \item[\textbf{$\beta$16}.] Open the right eye only, then fixate.
    \item[\textbf{$\beta$17}.] Open the left eye only, then fixate.
    \item[\textbf{$\beta$18}.] Clockwise sweep along a rectangular frame, indicating forward increase (e.g., volume up).
    \item[\textbf{$\beta$19}.] Counterclockwise sweep along a rectangular frame, indicating backward decrease (e.g., volume down).
    \item[\textbf{$\beta$20}.] Fixate on the top-right corner, then move to the bottom-right corner.
    \item[\textbf{$\beta$21}.] Perform a Z-shaped sweep: top-left $\rightarrow$ top-right $\rightarrow$ bottom-left $\rightarrow$ bottom-right.
    \item[\textbf{$\beta$22}.] Perform a V-shaped sweep: top-left $\rightarrow$ bottom-center $\rightarrow$ top-right, or the mirrored direction.
    \item[\textbf{$\beta$23}.] Blink $N$ times.
    \item[\textbf{$\beta$24}.] Converge both eyes inward (cross-eye gesture).
    \item[\textbf{$\beta$25}.] Follow a dot grid: top-left $\rightarrow$ bottom-right $\rightarrow$ top-right $\rightarrow$ bottom-left $\rightarrow$ top-left.
    \item[\textbf{$\beta$26}.] Widen eyes to indicate ``increase'', narrow eyes to indicate ``decrease''.
\end{enumerate}
}

\end{document}